\documentclass[12pt]{article}

\usepackage[margin=1in]{geometry}
\usepackage{setspace}
\usepackage{amsmath,amssymb,amsfonts}
\usepackage{booktabs}
\usepackage{graphicx}
\usepackage{natbib}
\usepackage{hyperref}
\usepackage{xcolor}
\usepackage{threeparttable}
\usepackage{subcaption}
\usepackage{lineno}

\hypersetup{
    colorlinks=true,
    linkcolor=blue,
    citecolor=blue,
    urlcolor=blue
}

\title{\textbf{The Shape of Macroeconomic Beliefs}}

\author{Giovanni Angelini\thanks{Corresponding author: Giovanni Angelini, Department of Economics, University of Bologna, Piazza Scaravilli 2, 40126 Bologna, Italy. E-mail: \href{mailto:g.angelini@unibo.it}{g.angelini@unibo.it}} \textsuperscript{a}\\[0.15cm]
\small \textsuperscript{a}University of Bologna}

\date{June 2026}

\begin{document}
\maketitle

\begin{abstract}
\noindent

\singlespacing
Macroeconomic expectations are usually observed through point forecasts or through asset prices whose mapping into beliefs is model-dependent. This paper uses prediction-market prices to recover high-frequency distributions of short-run macroeconomic beliefs. We construct a panel of Kalshi-implied distributions for CPI and core CPI releases by converting adjacent threshold contracts into probability mass over inflation outcomes. The data reveal market-implied means, uncertainty, and upper-tail probabilities from 30 days to one hour before each release. The market-implied mean contains meaningful forecast information, especially for headline CPI, but the main signal is distributional. Lagged Reuters Poll surprises do not predict systematic deviations of Kalshi means from the current Reuters consensus. By contrast, large lagged surprises are associated with higher implied uncertainty, and positive lagged surprises raise the probability assigned to fixed high-inflation outcomes. In the baseline specification with variable-by-horizon fixed effects, a 0.1 percentage point positive lagged surprise raises the probability of monthly inflation above 0.3 percent by about 4.7 percentage points, even after controlling for the current consensus forecast. In release-level validation tests, Kalshi upper-tail probabilities also predict the realization of high-inflation states, including episodes in which the market-implied mean remains close to the Reuters consensus. The evidence suggests that prediction markets can provide real-time information about inflation risk that is missed by point forecasts.

\vspace{0.2cm}


\end{abstract}

\clearpage

\section{Introduction}
\label{sec:introduction}

Expectations matter for monetary policy because decisions depend not only on the most likely future outcome, but also on the risks around that outcome. The distinction is especially important for inflation. A central bank faces a different communication problem when the expected CPI release is stable but the probability of a high-inflation outcome has risen. Yet many empirical measures of macroeconomic expectations are incomplete along this dimension. Surveys are direct, but are often low frequency or summarized by point forecasts. Survey densities are valuable but usually refer to lower-frequency horizons rather than to the probability distribution of a specific statistical release. Asset prices are high frequency, but extracting beliefs from them requires assumptions about risk premia, liquidity, and the mapping from asset payoffs to macroeconomic states. The object that is often most useful for monetary monitoring is a release-level subjective distribution; the econometrician usually observes only a mean or a model-dependent signal. The analysis uses prediction markets to study the distribution of short-run macroeconomic beliefs. A panel of market-implied distributions is constructed from macroeconomic event contracts traded on Kalshi. The key empirical feature is that many contracts are written on adjacent thresholds. For example, separate contracts may pay depending on whether monthly CPI inflation exceeds 0.2, 0.3, or 0.4 percent. Observing prices across these thresholds before the release makes it possible to recover a discrete distribution over the upcoming macroeconomic outcome. The resulting data reveal the market-implied mean, dispersion, skewness, entropy, and tail probabilities at standardized horizons from 30 days to one hour before resolution. The central question is whether recent macroeconomic news affects the location of prediction-market beliefs or their shape. A mean-based analysis asks whether a positive inflation surprise shifts the next expected inflation release upward. A distributional analysis asks a broader question: does the market reallocate probability mass toward high-inflation states, or become more uncertain, even when the mean remains close to consensus? This distinction matters for monetary economics. Prediction markets may be useful not because they always dominate professional forecasts in predicting the modal outcome, but because they reveal how markets price the tails of the next release.

The analysis focuses on CPI and core CPI month-over-month releases, which provide the cleanest repeated setting. The baseline distributional sample contains 282 CPI and 240 core CPI snapshots across 46 and 39 releases, respectively. These distributions are merged with Reuters Poll consensus forecasts. This allows the empirical analysis to separate three objects: the realized release, the professional consensus forecast, and the contemporaneous prediction-market distribution. The Reuters consensus is not treated as an exogenous object. It is a conditioning benchmark for the professional point forecast. The reduced-form question is whether prediction-market distributions contain additional information about residual inflation risk, conditional on that benchmark and on professional-forecaster disagreement. The first result is that Kalshi-implied means contain meaningful forecast information. For headline CPI, the Kalshi mean has an overall RMSE of 0.169 percentage points, compared with 0.331 for the previous realization, 0.280 for a three-month moving average, and 0.274 for a six-month moving average. For core CPI, Kalshi is broadly competitive with simple moving-average benchmarks, although the improvement is smaller. This forecast content is important because it shows that the recovered distributions are not noise. Forecast accuracy, however, is not the main contribution. The second result is that recent inflation news is not primarily reflected in disagreement about the center of the next release distribution. With variable-by-horizon fixed effects, the coefficient of the mean gap \(\mu^K-C\) on the lagged Reuters surprise is \(-0.023\), with a clustered standard error of 0.107. Thus, past Reuters surprises do not lead Kalshi means to deviate systematically from the current Reuters consensus. This null result is informative: the market's distinctive signal is not a persistent shift in the conditional mean relative to professional forecasters. The third and main result is distributional. Large previous surprises are associated with greater market-implied uncertainty. Positive previous surprises raise the probability assigned to fixed high-inflation outcomes. With variable-by-horizon fixed effects and the current Reuters consensus controlled for, the coefficient on the lagged Reuters surprise is 0.466 for \(\Pr^K(y>0.3)\), 0.422 for \(\Pr^K(y>0.4)\), and 0.281 for \(\Pr^K(y>0.5)\). These magnitudes imply that a 0.1 percentage point positive lagged surprise raises the probability of inflation above 0.3 percent by about 4.7 percentage points. The response is weaker when tails are defined relative to the contemporaneous consensus forecast. The interpretation is therefore not that Kalshi mechanically disagrees with Reuters after a surprise. Instead, recent inflation news changes the absolute inflation-risk environment: the mean remains close to consensus, while the market assigns more mass to high-inflation states.

Several validation and robustness exercises support this interpretation. Release-level tests show that Kalshi upper-tail probabilities predict the realization of high-inflation states, including episodes in which the market-implied mean remains close to the Reuters consensus. Staleness filters from 24 hours to one hour do not attenuate the uncertainty and fixed-tail coefficients. Direct controls for snapshot staleness, open interest, and volume leave the central estimates positive. Alternative tail-support rules and uniform-within-bin interpolation deliver the same qualitative fixed-tail result. Wild cluster bootstrap inference by release is more conservative, but the main uncertainty and fixed-tail estimates remain statistically meaningful. Split-sample regressions for CPI and core CPI preserve the signs, although precision declines. Leave-one-release-out diagnostics show that the uncertainty and fixed-tail estimates are not driven by a single inflation release.

The paper contributes to four literatures. First, it contributes to work on expectation formation and information frictions. Rational-expectations and noisy-information models emphasize the role of information sets, attention, and dispersed signals in shaping beliefs \citep{Muth1961,Lucas1972,MankiwReis2002,Sims2003,Woodford2003,MorrisShin2002,Carroll2003,CoibionGorodnichenko2012,CoibionGorodnichenko2015}. The contribution is financially incentivized, high-frequency evidence on the full distribution of short-run inflation beliefs. Second, it relates to behavioral models of extrapolation, diagnostic expectations, anchoring, and salience \citep{BarberisShleiferVishny1998,DanielHirshleiferSubrahmanyam1998,HongStein1999,BordaloGennaioliShleifer2012,BordaloGennaioliShleifer2018,BordaloGennaioliLaPortaShleifer2019}. The evidence is consistent with recent inflation news affecting uncertainty and tail probabilities more than one-for-one mean beliefs. Third, the paper contributes to the measurement of macroeconomic expectations, including work using survey densities \citep{Croushore1993,EngelbergManskiWilliams2009,AndradeCrumpEusepiMoench2016}. Prediction markets occupy an intermediate position between surveys and standard asset-price measures: their payoffs are directly tied to macroeconomic releases, but their prices are high frequency and financially backed. Fourth, the paper adds to the prediction-market literature, which studies information aggregation, calibration, risk preferences, and market microstructure \citep{ForsytheNelsonNeumannWright1992,BergNelsonRietz2008,WolfersZitzewitz2004,Manski2006,WolfersZitzewitz2006,ArrowEtAl2008,SnowbergWolfersZitzewitz2013}. Recent platform-specific work studies Kalshi and Polymarket in macroeconomic, political, and microstructure applications \citep{DiercksKatzWright2026,SwansonWangWu2025,BurgiDengWhelan2026,EichengreenViswanathNatrajWangWang2025,TsangYang2026,TsangYang2026PoliticalShocks,Dubach2026,AkeyGregoireHarvieMartineau2026,GomezCramGuoJensenKung2026Accuracy,GomezCramGuoJensenKung2025Earnings,SaguilloGhafouriKifferSuarezTangil2025,DudleyMagdaleno2026,ClintonHuang2025}. The analysis is complementary to that literature because it uses prediction-market prices as a measurement device for distributional inflation beliefs and studies how recent macroeconomic news affects the shape of those beliefs. The online appendix gives a detailed comparison with recent platform-specific papers.

The rest of the paper is organized as follows. Section~\ref{sec:data} describes the data and the construction of market-implied distributions. Section~\ref{sec:framework} presents the empirical framework. Section~\ref{sec:results} reports the main results on forecast content, consensus alignment, uncertainty, and tail risk. Section~\ref{sec:robustness} reports robustness and diagnostic checks. Section~\ref{sec:discussion} discusses interpretation, limitations, and policy relevance. Section~\ref{sec:conclusion} concludes.

\section{Data and Market-Implied Belief Distributions}
\label{sec:data}

This section describes the construction of the Kalshi distributional panel and the external forecast data used in the empirical analysis. The goal is to convert a cross-section of event-contract prices into a repeated panel of market-implied distributions for macroeconomic releases. The main analysis uses CPI and core CPI releases because these markets have the most regular contract structure and the closest mapping to Reuters Poll consensus forecasts.

\subsection{Kalshi macroeconomic contracts}

Kalshi is a real-money prediction-market exchange on which traders buy and sell contracts linked to the realization of well-defined events. In the macroeconomic markets studied here, the event is tied to a future statistical release, such as monthly CPI inflation, core CPI inflation, nonfarm payrolls, the unemployment rate, PCE inflation, Treasury rates, or recession-related outcomes. Contracts have binary payoffs. A contract that pays one dollar if an event occurs can be interpreted as a market price for that event, subject to the usual caveats about fees, risk preferences, liquidity, bid-ask spreads, market power, and heterogeneous beliefs \citep{Manski2006,WolfersZitzewitz2004,WolfersZitzewitz2006,BurgiDengWhelan2026}. The raw data contain contract identifiers, event identifiers, contract titles and subtitles, parsed threshold values, timestamps, prices, volume, and open interest when available. Prices are aggregated to an hourly frequency. Hourly observations preserve the high-frequency nature of the market while reducing the influence of isolated quote-level noise. For each event and hour, the cross-section of active contracts is retained and the implied distribution is recovered when there are enough contracts to identify at least three probability bins. Table~\ref{tab:coverage} summarizes the market coverage. Panel A reports the broad macroeconomic sample. The raw panel contains 476 macroeconomic events and more than 5,000 contracts, corresponding to more than 760,000 hourly contract-level observations. CPI and core CPI are the largest and cleanest repeated settings, with 100 events each. Panel B reports the baseline month-over-month inflation sample used in the main analysis. After imposing the distributional filters and standardized snapshot horizons, the sample contains 282 CPI and 240 core CPI distributional snapshots across 46 and 39 events. The median implied mean is about 0.25 percentage points month over month, and the median implied standard deviation is about 0.10 percentage points.

\begin{table}[!htbp]
\centering
\caption{Market coverage and baseline inflation sample}
\label{tab:coverage}
\begin{threeparttable}
\scriptsize
\resizebox{0.94\textwidth}{!}{%
\begin{tabular}{lccccccc}
\toprule
\multicolumn{8}{c}{\textbf{Broad macroeconomic prediction-market coverage}} \\
\midrule
Variable & Events & Contracts & Contract-hour obs. & Distribution obs. & Snapshot events & Snapshot obs. & Avg. bins \\
\midrule
CPI                    & 100 & 1,148 & 201,210 & 33,137 & 73 & 503 & 5.37 \\
Core CPI               & 100 & 1,116 & 122,800 & 23,930 & 77 & 528 & 4.90 \\
GDP / recession         &  30 &   221 &  90,550 & 11,148 & 13 &  89 & 6.16 \\
Nonfarm payrolls        &  47 &   444 & 143,480 & 17,909 & 36 & 245 & 5.70 \\
PCE inflation           &  28 &   146 &  17,344 &  5,061 & 17 & 106 & 4.08 \\
Treasury rates          & 100 & 1,375 &  47,637 &  4,228 & 83 & 251 & 8.11 \\
Unemployment       &  71 &   823 & 140,160 & 13,444 & 45 & 306 & 5.26 \\
\bottomrule
\end{tabular}}

\centering
\vspace{0.1cm}
\resizebox{0.94\textwidth}{!}{
\begin{tabular}{lccccccc}
\multicolumn{8}{c}{\textbf{Baseline CPI and core CPI month-over-month sample}} \\
\midrule
Variable group & Events & Snapshot obs. & Median bins & Median mean & Median s.d. & Mean stale hours & Max stale hours \\
\midrule
CPI      & 46 & 282 & 4.5 & 0.249 & 0.105 & 6.21 & 23.42 \\
Core CPI & 39 & 240 & 4.0 & 0.259 & 0.094 & 6.91 & 23.42 \\
\bottomrule
\end{tabular}}

\vspace{0.15cm}

\begin{minipage}{0.88\textwidth}
\footnotesize
\emph{Notes:} Top panel reports coverage of Kalshi macroeconomic prediction markets by variable group. Contract-hour observations are hourly contract-level observations. Distribution observations are event-time observations for which a market-implied distribution can be recovered from the available cross-section of contracts. Snapshot observations are distributions constructed at fixed horizons before resolution. Bottom panel reports the baseline CPI and core CPI month-over-month sample used in the main analysis. Market-implied means and standard deviations are measured in percentage points month over month. Stale hours measure the time between the target snapshot horizon and the most recent available market observation.
\end{minipage}
\end{threeparttable}
\end{table}

\subsection{Recovering distributions from threshold prices}

\begin{figure}[!htbp]
\centering
\includegraphics[width=0.72\textwidth]{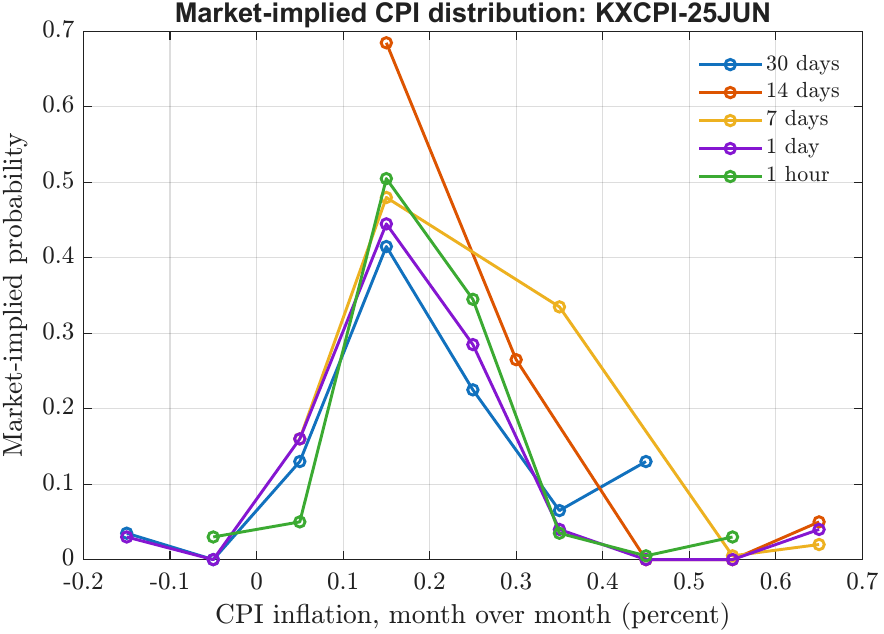}
\caption{Market-implied distribution of CPI beliefs before release}
\label{fig:distribution_example}
\begin{minipage}{0.88\textwidth}
\footnotesize
\emph{Notes:} The figure reports the Kalshi-implied probability distribution for a CPI month-over-month release at selected horizons before resolution. The distribution is recovered from threshold contracts and normalized across adjacent bins.
\end{minipage}
\end{figure}

Let \(y_{g,r}\) denote the outcome for variable group \(g\) and release \(r\). For inflation, \(y_{g,r}\) is the month-over-month CPI or core CPI inflation rate. At horizon \(h\), the Kalshi price cross-section implies a distribution
\[
F^K_{g,r,h}(y)=\Pr^K_{g,r,h}(y_{g,r}\leq y),
\]
where the superscript \(K\) denotes Kalshi. Many contracts in the sample are threshold contracts. If a contract pays one dollar when \(y_{g,r}>c_j\), its price is interpreted as an implied survival probability,
\[
p_{g,r,h}(c_j)\approx \Pr^K_{g,r,h}(y_{g,r}>c_j).
\]
For ordered thresholds \(c_1<\cdots<c_J\), adjacent prices identify probability mass over bins. For interior bins,
\[
\pi^K_{g,r,h,j}=\Pr^K_{g,r,h}(c_j<y_{g,r}\leq c_{j+1})=p_{g,r,h}(c_j)-p_{g,r,h}(c_{j+1}).
\]
The lower and upper open-ended bins are
\[
\pi^K_{g,r,h,0}=1-p_{g,r,h}(c_1),\qquad
\pi^K_{g,r,h,J}=p_{g,r,h}(c_J).
\]
When markets are quoted as bracket contracts rather than cumulative threshold contracts, the bracket prices are used directly as bin probabilities. In either case, small no-arbitrage violations can arise from stale quotes, bid-ask effects, or asynchronous trading. The baseline procedure truncates negative probability masses at zero and renormalizes the distribution to sum to one. The empirical analysis records the raw mass and the number of bins as diagnostics. Each bin receives a representative midpoint \(m_j\). Closed bins use the midpoint of the interval. Open-ended bins use a conservative midpoint based on the adjacent grid width. This choice affects the mean and standard deviation when substantial mass lies in the open tails. The online appendix therefore reports robustness exercises that vary tail supports and interpolate survival probabilities within bins. Given normalized bin probabilities \(\pi^K_{g,r,h,j}\), the market-implied mean is
\[
\mu^K_{g,r,h}=\sum_j \pi^K_{g,r,h,j}m_j,
\]
and the standard deviation is
\[
\sigma^K_{g,r,h}=\left[\sum_j \pi^K_{g,r,h,j}(m_j-\mu^K_{g,r,h})^2\right]^{1/2}.
\]
The recovered distribution also yields skewness, entropy, the maximum bin probability, and economically relevant tail probabilities,
\[
Tail^K_{g,r,h}(a)=\Pr^K_{g,r,h}(y_{g,r}>a).
\]
For inflation, the main tail thresholds are fixed monthly rates, such as 0.3, 0.4, and 0.5 percent, as well as thresholds relative to the Reuters consensus forecast. For each release, snapshots are constructed at seven standardized horizons before market resolution:
\[
h\in\{30\text{ days},14\text{ days},7\text{ days},3\text{ days},1\text{ day},6\text{ hours},1\text{ hour}\}.
\]
At each horizon, the most recent available distribution before the target time is selected and its staleness is recorded. This creates a release-by-horizon panel of distributional objects. The baseline CPI analysis uses snapshots with enough active bins to recover a meaningful distribution. Figure~\ref{fig:distribution_example} illustrates the recovered object for one CPI release.

\subsection{Consensus forecasts and realized releases}

\begin{figure}[!htbp]
\centering
\begin{subfigure}{0.48\textwidth}
\centering
\includegraphics[width=\textwidth]{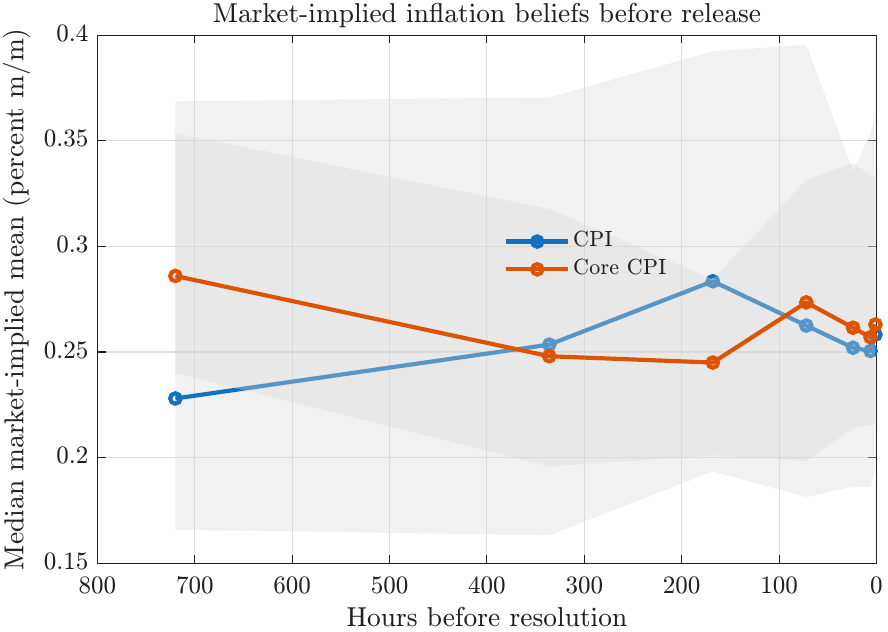}
\caption{Market-implied mean}
\end{subfigure}
\hfill
\begin{subfigure}{0.48\textwidth}
\centering
\includegraphics[width=\textwidth]{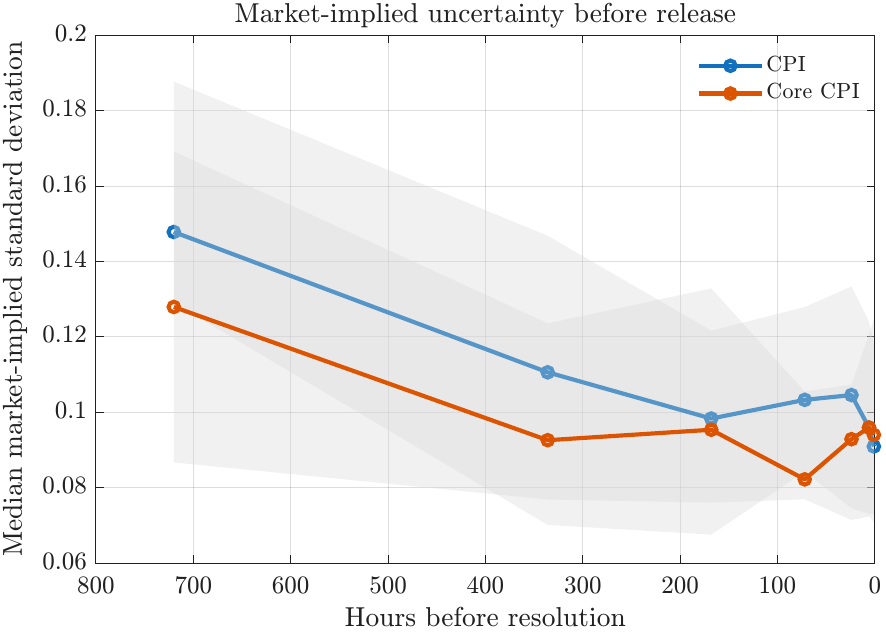}
\caption{Market-implied uncertainty}
\end{subfigure}
\caption{Market-implied inflation beliefs before release}
\label{fig:mean_uncertainty}
\begin{minipage}{0.88\textwidth}
\footnotesize
\emph{Notes:} Panel A reports the median market-implied mean for CPI and core CPI month-over-month releases at different horizons before resolution. Panel B reports the corresponding median market-implied standard deviation. Shaded areas denote interquartile ranges across releases.
\end{minipage}
\end{figure}

The main empirical benchmark is the Reuters Poll consensus forecast. For each CPI and core CPI release, the Reuters data report the median forecast, low forecast, high forecast, realized release, and the surprise defined as actual minus the median forecast. Let \(C_{g,r}\) denote the Reuters median forecast. The surprise is
\[
s_{g,r}=y_{g,r}-C_{g,r}.
\]
The Reuters high-low range is used as a measure of professional-forecaster disagreement. CPI and core CPI realized values are also constructed from FRED indexes as \(100\times(I_t/I_{t-1}-1)\). The FRED-based actuals closely match the Reuters actuals in the merged sample; the mean difference is 0.001 percentage points for CPI and 0.004 percentage points for core CPI. This diagnostic is important because the analysis requires aligning the Kalshi event, the reference month, the Reuters poll, and the realized release. Figure~\ref{fig:mean_uncertainty} summarizes the recovered distributions across releases. The market-implied mean is economically plausible and relatively stable across horizons. Implied uncertainty is higher far from release and tends to decline as resolution approaches. This pattern is consistent with information arriving over time and being incorporated into the distribution of market beliefs.

\section{Empirical Framework}
\label{sec:framework}

The empirical tests are organized around a simple reduced-form framework in which the current professional consensus summarizes the center of the next release distribution. The framework is not a structural model of prediction-market trading. Its purpose is to clarify why recent inflation news may change the shape of market-implied beliefs even when the market-implied mean remains close to the professional consensus.

\subsection{Distributional updating}

Consider release \(r\) for variable \(g\), and let \(C_{g,r}\) denote the Reuters median consensus forecast. Write the release outcome as
\begin{equation}
    y_{g,r}=C_{g,r}+x_{g,r},
    \label{eq:residual_outcome}
\end{equation}
where \(x_{g,r}\) is the residual outcome relative to the contemporaneous professional point forecast. The prediction market prices a distribution over \(x_{g,r}\). Conditional on the market information set at horizon \(h\), suppose that this residual distribution can be represented as a mixture
\begin{equation}
    F^{K}_{g,r,h}(x)
    =
    \left(1-q_{g,r,h}\right)F^{0}_{g,h}(x;v_{g,r,h})
    +
    q_{g,r,h}F^{H}_{g,h}(x;v_{g,r,h}),
    \label{eq:mixture_distribution}
\end{equation}
where \(F^{0}\) is a normal-risk residual distribution, \(F^{H}\) is a high-inflation-risk residual distribution, \(q_{g,r,h}\in[0,1]\) is the market-implied probability of the high-risk state, and \(v_{g,r,h}\) governs residual uncertainty. The high-risk state need not be interpreted as a separate structural regime. It is a reduced-form way to capture the possibility that, after recent inflation news, market participants assign more probability to states in which the next release lies in the upper part of the inflation distribution. The key assumption is that the professional consensus absorbs much of the information about the center of the next release distribution. The prediction-market mean relative to the consensus is
\begin{equation}
    \mu^{K}_{g,r,h}-C_{g,r}
    =
    \int x\,dF^{K}_{g,r,h}(x).
    \label{eq:mixture_mean_gap}
\end{equation}
If the consensus forecast already incorporates the central implications of public information, or if the high-risk state mainly changes skewness and tail thickness rather than the center of the residual distribution, then changes in \(q_{g,r,h}\) need not generate a large movement in \(\mu^{K}_{g,r,h}-C_{g,r}\). In that case, the market-implied mean may remain close to the professional consensus even though the distribution around that mean changes. The same change in \(q_{g,r,h}\), however, has direct implications for upper-tail probabilities. For a fixed inflation threshold \(a\), the market-implied probability of a high-inflation outcome is
\begin{equation}
    \Pr^{K}_{g,r,h}(y_{g,r}>a)
    =
    1-F^{K}_{g,r,h}(a-C_{g,r}).
    \label{eq:mixture_tail}
\end{equation}
Holding \(C_{g,r}\) fixed, the effect of the high-risk weight on this probability is
\begin{equation}
    \frac{\partial \Pr^{K}_{g,r,h}(y_{g,r}>a)}
         {\partial q_{g,r,h}}
    =
    F^{0}_{g,h}(a-C_{g,r};v_{g,r,h})
    -
    F^{H}_{g,h}(a-C_{g,r};v_{g,r,h}).
    \label{eq:mixture_tail_derivative}
\end{equation}
For high absolute inflation thresholds, such as 0.3, 0.4, or 0.5 percent month over month, the high-risk distribution assigns more mass above the threshold than the normal-risk distribution. Thus \(F^{H}_{g,h}(a-C_{g,r};v_{g,r,h})<F^{0}_{g,h}(a-C_{g,r};v_{g,r,h})\), and an increase in \(q_{g,r,h}\) raises the fixed upper-tail probability. The variance of the market-implied distribution also increases when either the high-risk state is more dispersed or the residual uncertainty parameter \(v_{g,r,h}\) rises. Recent macroeconomic news affects the market-implied distribution through these two objects. Let the previous Reuters surprise be \(s_{g,r-1}=y_{g,r-1}-C_{g,r-1}\). The high-inflation-risk weight is allowed to satisfy
\begin{equation}
    q_{g,r,h}
    =
    \Lambda\left(\alpha_{g,h}+\rho s_{g,r-1}+Z'_{g,r}\pi\right),
    \qquad \rho>0,
    \label{eq:q_index}
\end{equation}
where \(\Lambda(\cdot)\) maps the index into \([0,1]\), and \(Z_{g,r}\) contains contemporaneous controls such as the current consensus forecast. A positive lagged inflation surprise therefore raises the probability attached to the high-inflation-risk state. Residual uncertainty may instead respond to the size of recent news:
\begin{equation}
    v_{g,r,h}
    =
    \nu_{g,h}
    +
    \eta |s_{g,r-1}|
    +
    \theta Range^{R}_{g,r},
    \qquad \eta>0,
    \label{eq:v_index}
\end{equation}
where \(Range^{R}_{g,r}\) is the Reuters high-low forecast range. This captures the idea that large recent surprises make the next release more uncertain, even after controlling for contemporaneous disagreement among professional forecasters. The framework delivers four empirical implications. First, the response of the market-implied mean relative to the Reuters consensus, \(\mu^{K}_{g,r,h}-C_{g,r}\), may be small even when recent news matters. Second, large lagged surprises should be associated with higher market-implied uncertainty. Third, positive lagged surprises should raise fixed upper-tail probabilities such as \(\Pr^{K}_{g,r,h}(y_{g,r}>0.3)\), \(\Pr^{K}_{g,r,h}(y_{g,r}>0.4)\), and \(\Pr^{K}_{g,r,h}(y_{g,r}>0.5)\), conditional on the current consensus. Fourth, the response of tails defined relative to the current consensus, \(\Pr^{K}_{g,r,h}(y_{g,r}>C_{g,r}+q)\), may be weaker because these thresholds move with the professional forecast and therefore remove part of the absolute inflation-risk component. The interpretation is deliberately reduced form. The Reuters consensus is a conditioning benchmark, not an exogenous treatment. The regressions below should therefore be read as predictive conditional relationships, not as causal estimates of belief updating. Their purpose is to test whether recent inflation news is reflected mainly in the location of market-implied beliefs or instead in uncertainty and upper-tail probabilities.

\subsection{Location}

The first test asks whether recent inflation surprises shift prediction-market means away from the contemporaneous professional consensus. The estimating equation is
\begin{equation}
\mu^K_{g,r,h}-C_{g,r}
=
\alpha_{g,h}+\beta s_{g,r-1}+\varepsilon_{g,r,h},
\label{eq:mean_gap}
\end{equation}
where \(\alpha_{g,h}\) denotes variable-by-horizon fixed effects. A positive \(\beta\) would indicate extrapolation in the central forecast: after an above-consensus release, the prediction-market mean for the next release is higher than the current Reuters consensus. A coefficient close to zero does not imply that the market ignores recent news. It means only that recent news does not create a systematic gap between the Kalshi mean and the professional consensus.

\subsection{Uncertainty}

The second test studies whether large surprises widen the market-implied distribution. The specification is
\begin{equation}
\sigma^K_{g,r,h}
=
\alpha_{g,h}+\eta |s_{g,r-1}|+\theta Range^R_{g,r}+u_{g,r,h},
\label{eq:uncertainty}
\end{equation}
where \(Range^R_{g,r}\) is the Reuters high-low forecast range for the current release. The range control is useful because professional-forecaster disagreement may itself be high when the current release is difficult to forecast. The coefficient \(\eta\) therefore asks whether the market-implied distribution is wider after large previous surprises, conditional on contemporaneous survey disagreement.

\subsection{Tail risk}

The third test examines upper-tail probabilities. For fixed inflation thresholds \(a=0.3,0.4,0.5\), the specification is
\begin{equation}
Tail^K_{g,r,h}(a)
=
\alpha_{g,h}+\delta s_{g,r-1}+\gamma C_{g,r}+v_{g,r,h}.
\label{eq:fixed_tail}
\end{equation}
The current consensus control matters because a higher expected current release mechanically raises the probability of exceeding a fixed high-inflation threshold. A positive \(\delta\) in equation~\eqref{eq:fixed_tail} means that recent inflation news predicts additional probability mass in high-inflation states, beyond what is explained by the current consensus forecast. Additional tail specifications let the threshold move with the current consensus,
\begin{equation}
\Pr^K_{g,r,h}(y_{g,r}>C_{g,r}+q)
=
\alpha_{g,h}+\delta_q s_{g,r-1}+\theta_q Range^R_{g,r}+e_{g,r,h},
\label{eq:relative_tail}
\end{equation}
for \(q\in\{0,0.1\}\). These relative-tail regressions distinguish a general increase in absolute inflation risk from a systematic disagreement with the current consensus. If lagged surprises raise fixed high-inflation tails but not consensus-relative tails, the interpretation is that the inflation-risk environment has changed, not that prediction markets are simply biased above the current professional forecast. All baseline standard errors are clustered by release event, which accounts for dependence across horizons within the same statistical release. The main specifications pool CPI and core CPI and include variable-by-horizon fixed effects; the tables also report simpler horizon and variable fixed-effect specifications where useful. The online appendix reports moving-average benchmark exercises and labor-market external-validity tests.

\section{Results}
\label{sec:results}

This section reports the main empirical results. The central finding is that the Kalshi-implied mean is informative, but it is not the main margin on which recent inflation news appears. Lagged Reuters surprises do not predict systematic movements of Kalshi means away from the current Reuters consensus. Instead, recent surprises are reflected in distributional shape: uncertainty and fixed high-inflation tail probabilities.

\subsection{Forecast content of the market-implied mean}

The first exercise evaluates whether the recovered market-implied means contain meaningful information about the upcoming release. For each release \(r\) and horizon \(h\), the Kalshi-implied mean \(\mu^K_{r,h}\) is compared with realized month-over-month inflation \(y_r\). The benchmark forecasts are the previous realization, a three-month moving average, and a six-month moving average, all constructed using only past realizations. Table~\ref{tab:forecast_performance} reports the results.

\begin{table}[!htbp]
\centering
\caption{Forecast performance of market-implied inflation means}
\label{tab:forecast_performance}
\begin{threeparttable}
\small
\begin{tabular}{lcccccccc}
\toprule
Variable & Obs. & Events & Bias & MAE & RMSE & RMSE Prev. & RMSE MA3 & RMSE MA6 \\
\midrule
CPI      & 268 & 44 & 0.049 & 0.117 & 0.169 & 0.331 & 0.280 & 0.274 \\
Core CPI & 227 & 37 & 0.026 & 0.094 & 0.118 & 0.122 & 0.114 & 0.111 \\
\bottomrule
\end{tabular}
\begin{tablenotes}
\footnotesize
\item \emph{Notes:} The table compares realized month-over-month inflation with the Kalshi-implied mean and simple benchmark forecasts. Bias is realized inflation minus the Kalshi-implied mean. MAE and RMSE are measured in percentage points month over month. ``Prev.'' denotes the previous monthly realization. MA3 and MA6 denote three- and six-month moving averages constructed using only past realizations.
\end{tablenotes}
\end{threeparttable}
\end{table}

\begin{figure}[!htbp]
\centering
\includegraphics[width=0.78\textwidth]{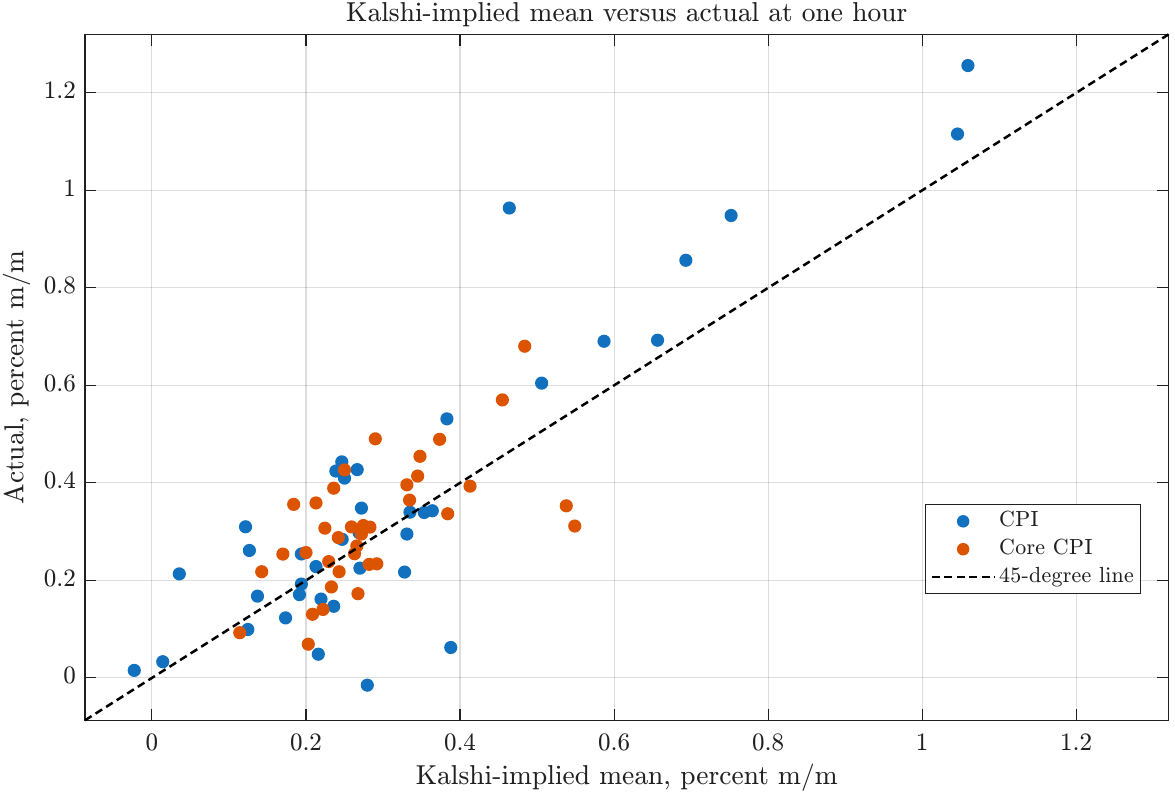}
\caption{Kalshi-implied mean versus realized inflation at one hour}
\label{fig:kalshi_actual}
\begin{minipage}{0.88\textwidth}
\footnotesize
\emph{Notes:} The figure plots realized month-over-month inflation against the Kalshi-implied mean at the one-hour horizon. The dashed line is the 45-degree line. Units are percentage points month over month.
\end{minipage}
\end{figure}

For headline CPI, the Kalshi mean substantially improves on simple time-series benchmarks. Its RMSE is 0.169 percentage points, compared with 0.331 for the previous realization, 0.280 for the three-month moving average, and 0.274 for the six-month moving average. For core CPI, the improvement is more modest: Kalshi is broadly competitive with rolling-average benchmarks but does not dominate them. Figure~\ref{fig:kalshi_actual} plots the one-hour-ahead Kalshi-implied mean against the realized release. The figure shows substantial information in the market mean, while also making clear that individual events can contain large errors.

\subsection{Consensus alignment and distributional updating}

The Kalshi panel is next merged with Reuters Poll consensus forecasts. For each CPI and core CPI release, the Reuters data report the median forecast, low forecast, high forecast, actual release, and surprise. The merged estimation sample contains 484 release-horizon observations with lagged Reuters surprises: 262 for CPI and 222 for core CPI. These observations correspond to 79 release clusters. Standard errors are therefore clustered by release throughout the baseline analysis, and release-level validation exercises below provide a check that the results are not generated by treating horizons as independent events.

The Reuters actuals closely match the FRED-based actuals used in the Kalshi panel. The mean difference between the FRED-based actual and the Reuters actual is 0.001 percentage points for CPI and 0.004 percentage points for core CPI, supporting the event-date alignment. Table~\ref{tab:reuters_distributional_dynamics} summarizes the main regressions. Panel A begins with the mean gap \(\mu^K-C\). With variable-by-horizon fixed effects, the coefficient on the lagged Reuters surprise is \(-0.023\), with a clustered standard error of 0.107. The estimate is economically small and statistically indistinguishable from zero. Past inflation surprises therefore do not lead prediction-market means to deviate systematically from current Reuters consensus forecasts. The second row of Panel A studies uncertainty. The dependent variable is the Kalshi-implied standard deviation, and the regressor is the absolute lagged Reuters surprise. The specification controls for the Reuters high-low forecast range. The coefficient is 0.089, with a clustered standard error of 0.053. This estimate is less precise than the tail estimates below, but it is positive and economically consistent with the view that large recent surprises widen the market-implied distribution. The positive role of the Reuters range in the underlying horse-race regressions also validates the distributional measure: Kalshi-implied uncertainty comoves with professional-forecaster disagreement.

\begin{table}[!htbp]
\centering
\caption{Reuters surprises and market-implied belief distributions}
\label{tab:reuters_distributional_dynamics}
\begin{threeparttable}
\small
\begin{tabular}{lccccc}
\toprule
Dependent variable & Main regressor & Control & Coef. & s.e. & \(p\)-value \\
\midrule
\multicolumn{6}{c}{\textbf{Location and uncertainty}} \\
\(\mu^K-C\) 
& \(s_{r-1}\) 
& -- 
& -0.023 & 0.107 & 0.830 \\

\(\sigma^K\) 
& \(|s_{r-1}|\) 
& Reuters range 
& 0.089 & 0.053 & 0.094 \\

\addlinespace
\multicolumn{6}{c}{\textbf{Fixed high-inflation tails}} \\
\(\Pr^K(y>0.3)\) 
& \(s_{r-1}\) 
& Reuters consensus 
& 0.466 & 0.225 & 0.042 \\

\(\Pr^K(y>0.4)\) 
& \(s_{r-1}\) 
& Reuters consensus 
& 0.422 & 0.178 & 0.020 \\

\(\Pr^K(y>0.5)\) 
& \(s_{r-1}\) 
& Reuters consensus 
& 0.281 & 0.134 & 0.039 \\

\addlinespace
\multicolumn{6}{c}{\textbf{Tails relative to consensus}} \\
\(\Pr^K(y>C)\) 
& \(s_{r-1}\) 
& Reuters range 
& 0.061 & 0.228 & 0.790 \\

\(\Pr^K(y>C+0.1)\) 
& \(s_{r-1}\) 
& Reuters range 
& 0.173 & 0.191 & 0.368 \\
\bottomrule
\end{tabular}
\begin{tablenotes}
\footnotesize
\item \emph{Notes:} The table reports regressions of Kalshi-implied belief-distribution objects on Reuters Poll surprises. \(C\) denotes the Reuters median consensus forecast and \(s_{r-1}\) is the previous release surprise, defined as actual minus Reuters median forecast. Reuters range is the difference between the Reuters high and low forecasts. All specifications include variable-by-horizon fixed effects. Standard errors are clustered by release. Inflation rates and surprises are measured in percentage points month over month.
\end{tablenotes}
\end{threeparttable}
\end{table}

\subsection{High-inflation tail risk}

\begin{figure}[!htbp]
\centering
\includegraphics[width=0.80\textwidth]{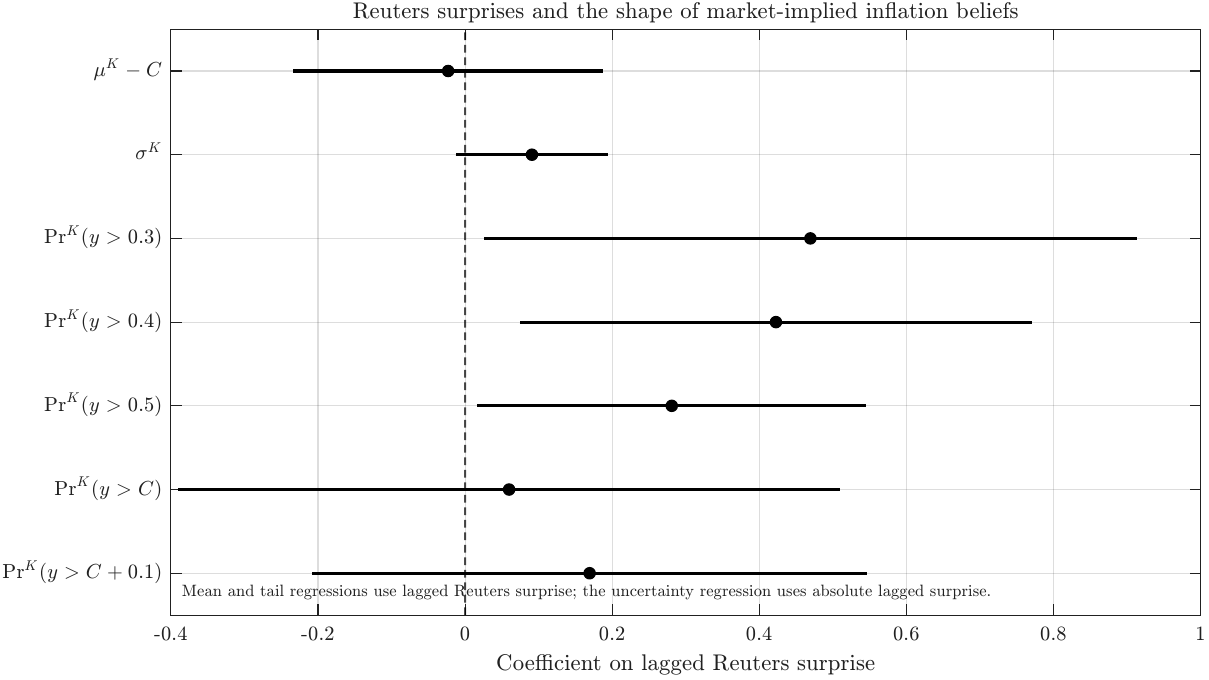}
\caption{Reuters surprises and the shape of market-implied inflation beliefs}
\label{fig:reuters_coefficients}
\begin{minipage}{0.9\textwidth}
\footnotesize
\emph{Notes:} The figure reports coefficients and 95 percent confidence intervals from regressions of Kalshi-implied belief-distribution objects on lagged Reuters Poll surprises. The mean regression uses \(\mu^K-C\) as the dependent variable. The uncertainty regression uses \(\sigma^K\) and the absolute lagged surprise. Tail regressions use fixed high-inflation thresholds and tails relative to the Reuters consensus forecast. All specifications include variable-by-horizon fixed effects. Standard errors are clustered by release.
\end{minipage}
\end{figure}

The strongest evidence appears in upper-tail probabilities. Panel B of Table~\ref{tab:reuters_distributional_dynamics} reports regressions for \(\Pr^K(y>0.3)\), \(\Pr^K(y>0.4)\), and \(\Pr^K(y>0.5)\), controlling for the current Reuters consensus forecast and variable-by-horizon fixed effects. The coefficients on the lagged Reuters surprise are 0.466, 0.422, and 0.281, respectively. The magnitudes are large. A 0.1 percentage point positive surprise in the previous release raises the probability assigned to inflation above 0.3 percent by about 4.7 percentage points, even after conditioning on the current consensus forecast. Panel C shows that the result is weaker when tails are defined relative to the current consensus forecast. The coefficient on \(\Pr^K(y>C)\) is 0.061 and statistically insignificant; the coefficient on \(\Pr^K(y>C+0.1)\) is 0.173 and also statistically insignificant. This difference is informative. Lagged surprises do not appear to make Kalshi systematically disagree with Reuters about whether the current release will exceed consensus. Instead, positive previous surprises raise the probability of fixed high-inflation outcomes. Recent inflation news therefore changes the perceived absolute inflation-risk environment. Figure~\ref{fig:reuters_coefficients} summarizes the same evidence graphically. The coefficient for the mean gap is close to zero. The uncertainty and fixed-tail coefficients are positive. The consensus-relative tail coefficients are small and imprecise. This is the central empirical message: point forecasts miss the main signal in prediction-market data.

\subsection{Inflation-at-risk as an early-warning measure}
\label{subsec:early_warning}

The previous results show that recent inflation news is reflected in the shape of the market-implied distribution rather than in a systematic gap between the Kalshi mean and the Reuters consensus. A natural question is whether these upper-tail probabilities are empirically meaningful as measures of inflation risk. This subsection therefore asks whether Kalshi-implied tail probabilities forecast the realization of high-inflation states. For each release-horizon observation, define realized tail indicators
\[
    \mathbf{1}\{y_{g,r}>a\},
    \qquad
    a\in\{0.3,0.4,0.5\},
\]
and compare them with the corresponding Kalshi-implied probabilities \(\Pr^K_{g,r,h}(y_{g,r}>a)\).

The exercise is deliberately simple. It does not ask whether prediction markets dominate every survey-based benchmark in every scoring rule. Instead, it asks whether the market-implied upper tail has useful early-warning content for adverse inflation outcomes, especially in states where the mean forecast may remain close to consensus. Table~\ref{tab:early_warning_tailrisk} reports two validation exercises. Panel A sorts release-horizon observations into terciles of the Kalshi-implied probability of inflation above 0.4 percent and compares the mean predicted probability with the realized frequency of the event. The realized frequency rises sharply across the tail-risk distribution. At the one-hour horizon, the event \(y>0.4\) occurs in 7.7 percent of low-tail observations, 12.0 percent of middle-tail observations, and 57.7 percent of high-tail observations. The corresponding mean Kalshi probabilities are 0.0, 8.0, and 60.5 percent. Panel B reports linear probability regressions of realized high-inflation indicators on the corresponding Kalshi-implied tail probability. The regressions include Reuters controls: the current Reuters median consensus and the Reuters high-low forecast range. At the one-hour horizon, the coefficient on \(\Pr^K(y>0.3)\) is 0.744, with a standard error of 0.159. For the more extreme event \(y>0.5\), the coefficient is 0.722, with a standard error of 0.272. The discrimination statistics are also strong: the AUC is 0.838 for \(y>0.3\), 0.815 for \(y>0.4\), and 0.926 for \(y>0.5\). These results provide a direct validation of the inflation-at-risk interpretation. Kalshi tail probabilities are not merely transformations of noisy contract prices. They identify states in which high-inflation outcomes are more likely to occur.

The results should not be read as showing that Kalshi probabilities uniformly dominate all survey-based distributional benchmarks. Rather, they show that prediction-market tails contain economically meaningful information about high-inflation states, including after conditioning on Reuters consensus information. This validation is also useful for interpretation. The main regressions show that lagged inflation surprises raise fixed upper-tail probabilities but do not systematically move the Kalshi mean away from Reuters consensus. The early-warning evidence shows why that margin matters: a stable central forecast can coexist with a market-implied distribution that assigns substantial probability to adverse inflation outcomes. In the one-hour release-level sample, there are 30 episodes in which the Kalshi mean is within 0.05 percentage points of the Reuters consensus but \(\Pr^K(y>0.4)\) lies in the upper tail-risk tercile. In these anchored-mean/high-tail episodes, the event \(y>0.4\) occurs in 50 percent of cases. This is the sense in which prediction-market distributions can provide a real-time measure of inflation-at-risk that is not visible in point forecasts alone.

\begin{table}[h!]
\centering
\caption{Kalshi upper-tail probabilities as early-warning indicators}
\label{tab:early_warning_tailrisk}
\begin{threeparttable}
\scriptsize
\resizebox{\textwidth}{!}{%
\begin{tabular}{llcccc}
\toprule
\multicolumn{6}{c}{\textbf{Calibration of $\Pr^K(y>0.4)$ by tail-risk tercile}} \\
\midrule
Horizon & Tail-risk bin & Obs. & Mean $\Pr^K(y>0.4)$ & Actual frequency $1\{y>0.4\}$ & Brier score \\
\midrule
7d & Low    & 25 & 0.000 & 0.200 & 0.200 \\
7d & Middle & 24 & 0.045 & 0.042 & 0.038 \\
7d & High   & 25 & 0.555 & 0.520 & 0.137 \\
1d & Low    & 25 & 0.000 & 0.160 & 0.160 \\
1d & Middle & 25 & 0.053 & 0.120 & 0.106 \\
1d & High   & 25 & 0.549 & 0.520 & 0.215 \\
1h & Low    & 26 & 0.000 & 0.077 & 0.077 \\
1h & Middle & 25 & 0.080 & 0.120 & 0.108 \\
1h & High   & 26 & 0.605 & 0.577 & 0.235 \\
\midrule
\multicolumn{6}{c}{\textbf{Predictive content at the one-hour horizon}} \\
\midrule
Event & Obs. & Actual frequency & AUC & Brier: Kalshi & LPM coefficient \\
\midrule
$1\{y>0.3\}$ & 77 & 0.506 & 0.838 & 0.175 & 0.744 \; (0.159) \\
$1\{y>0.4\}$ & 77 & 0.260 & 0.815 & 0.140 & 0.362 \; (0.192) \\
$1\{y>0.5\}$ & 77 & 0.143 & 0.926 & 0.045 & 0.722 \; (0.272) \\
\bottomrule
\end{tabular}}
\vspace{0.4em}
\begin{minipage}{0.94\textwidth}
\footnotesize
\emph{Notes:} Top panel sorts pooled CPI and core CPI release-horizon observations into terciles of the Kalshi-implied probability of inflation above 0.4 percent. The table reports the average predicted probability and the realized event frequency within each tercile. Bottom panel reports one-hour release-level validation statistics. AUC is the area under the ROC curve. The Brier score is the mean squared error of the Kalshi-implied tail probability for the corresponding realized event. The LPM coefficient is from a linear probability regression of the realized tail indicator on the corresponding Kalshi-implied tail probability, controlling for the Reuters median consensus and the Reuters high-low forecast range. Standard errors are in parentheses. Inflation rates are month-over-month percentage points.
\end{minipage}
\end{threeparttable}
\end{table}

\subsection{A monitoring exercise: point forecasts and inflation-at-risk}
\label{subsec:monitoring_exercise}

The previous subsection shows that Kalshi upper-tail probabilities contain early-warning information about high-inflation states. This subsection asks a more operational question: would a policymaker or market participant who monitors the upper tail of the Kalshi distribution identify high-inflation releases that are missed by point forecasts? I focus on headline CPI at the one-hour horizon and define a high-inflation release as
\[
    1\{y_r>0.4\},
\]
where \(y_r\) is monthly CPI inflation in percentage points. I compare three simple monitoring rules. The first signals high inflation when the Reuters median consensus exceeds \(0.4\) percent. The second signals high inflation when the Kalshi-implied mean exceeds \(0.4\) percent. The third uses the distributional object directly and signals high inflation when
\[
    \Pr^K_{r,-1h}(y_r>0.4) \geq 0.50 .
\]
The threshold of 50 percent is intentionally simple: it asks whether the market-implied distribution assigns the high-inflation state at least even odds. The exercise is not designed to optimize a classification rule. Its purpose is to illustrate the information lost when the monitoring problem is reduced to a point forecast.

Table~\ref{tab:false_comfort_monitoring} reports the results. Panel A first sorts releases into terciles of the Kalshi-implied probability of inflation above \(0.4\) percent. The realized frequency of high-inflation outcomes rises from 7.1 percent in the low-tail-risk tercile to 76.9 percent in the high-tail-risk tercile. The corresponding mean Kalshi probabilities are 0.4
and 74.9 percent, respectively. Thus, in this release-level monitoring sample, the upper tail of the Kalshi distribution ranks high-inflation risk in a way that is both monotone and economically large. Panel B compares the three monitoring rules. The Reuters consensus and the Kalshi-implied mean each identify 8 of the 13 high-inflation releases, with no false alarms. The Kalshi tail rule identifies 10 of the 13 high-inflation
releases, while generating only one false alarm. Its hit rate is therefore 76.9 percent, compared with 61.5 percent for either point-forecast rule, and its precision remains high at 90.9 percent. The overall accuracy of the tail rule is 90.0 percent. The exercise highlights the practical value of observing a distribution rather than only a mean. Point forecasts are conservative signals: when they cross a high-inflation threshold, the signal is very precise, but they miss several high-inflation outcomes. The Kalshi upper tail provides a complementary
monitoring statistic. It identifies additional high-inflation releases by using information about the shape of the market-implied distribution, not only its center. This is the sense in which prediction-market distributions can reduce the false comfort created by apparently stable point forecasts.

\begin{table}[h!]
\centering
\caption{Point forecasts and inflation-at-risk as monitoring signals}
\label{tab:false_comfort_monitoring}
\begin{tabular}{lcccc}
\toprule
\multicolumn{5}{c}{\textbf{Calibration by Kalshi upper-tail tercile}} \\
\midrule
Tail-risk bin $(y>0.4)$ & Obs. & Mean \(\Pr^K\) & Actual freq. &  \\
\midrule
Low    & 14 & 0.004 & 0.071 &  \\
Middle & 13 & 0.092 & 0.154 &  \\
High   & 13 & 0.749 & 0.769 &  \\
\midrule
\multicolumn{5}{c}{\textbf{Monitoring rules for high-inflation releases}} \\
\midrule
Signal & Hit rate & False alarm rate & Precision & Accuracy \\
\midrule
Reuters consensus \(>0.4\)             & 0.615 & 0.000 & 1.000 & 0.875 \\
Kalshi mean \(>0.4\)                   & 0.615 & 0.000 & 1.000 & 0.875 \\
Kalshi tail \(\Pr^K(y>0.4)\geq 0.50\)  & 0.769 & 0.037 & 0.909 & 0.900 \\
\bottomrule
\end{tabular}
\begin{minipage}{0.92\textwidth}
\footnotesize
\emph{Notes:} The table uses the headline CPI one-hour release-level monitoring sample. A high-inflation release is defined as monthly CPI inflation above 0.4 percent. Top panel sorts releases into terciles of the Kalshi-implied probability of \(y>0.4\). Panel B compares simple monitoring rules based on the Reuters median consensus, the Kalshi-implied mean, and the Kalshi-implied upper-tail probability. The hit rate is the share of high-inflation releases correctly signaled. The false alarm rate is the share of non-high-inflation releases incorrectly signaled. Precision is the share of signals that are followed by a high-inflation release. Accuracy is the share of all releases
correctly classified.
\end{minipage}
\end{table}

Section~\ref{sec:robustness} shows that this pattern is robust to quote staleness, liquidity controls, alternative tail-support assumptions, wild cluster bootstrap inference, CPI/core CPI splits, and leave-one-release-out diagnostics. These checks are important because the empirical object is a distribution recovered from a new and sometimes thinly traded market. They also sharpen the interpretation: the stable result is distributional risk, not a persistent shift in the conditional mean relative to Reuters consensus. Taken together, the results show that prediction-market beliefs are informative, but their distinctive contribution is distributional. Kalshi means contain information about realized inflation and remain aligned with professional consensus forecasts after past surprises. At the same time, recent inflation surprises reshape the market-implied distribution. Large surprises are associated with higher uncertainty, and positive surprises increase the probability of high-inflation states. This distributional margin is difficult to detect with point forecasts alone.

\section{Robustness and Diagnostics}
\label{sec:robustness}

The main results rely on a new market and on distributions recovered from discrete event-contract grids. This section summarizes robustness exercises designed to address the most direct threats to interpretation: stale quotes, liquidity, tail construction, finite-cluster inference, heterogeneity between CPI and core CPI, and influential releases. The tests support a narrow interpretation of the results. The robust finding is not a systematic drift of the market-implied mean away from professional consensus. It is that recent inflation news is reflected in distributional risk: implied uncertainty and fixed high-inflation tail probabilities.

Table~\ref{tab:robustness_summary} summarizes the main diagnostics. Staleness is not driving the results. The baseline panel has a median staleness of about 6.4 hours, and all observations are within 24 hours of the target snapshot. When the sample is restricted to snapshots no more than one hour stale, the implied-uncertainty coefficient remains positive and statistically significant, and the fixed-tail coefficients become larger. For example, the coefficient for \(\Pr^K(y>0.3)\) rises from 0.466 in the baseline to 1.198 in the one-hour-staleness subsample. The mean-gap coefficient remains imprecise. The results are also robust to controlling directly for market activity. Adding continuous controls for snapshot staleness, \(\log(1+\mathrm{open\ interest})\), and \(\log(1+\mathrm{volume})\) leaves the implied-standard-deviation coefficient positive. The coefficient for \(\Pr^K(y>0.4)\) remains significant at the five percent level, while the coefficients for \(\Pr^K(y>0.3)\) and \(\Pr^K(y>0.5)\) remain positive and marginally significant. Liquidity-filter results in the online appendix show a similar pattern when liquidity is measured by open interest. Volume filters are less stable because many hourly snapshots have zero measured volume even when open interest is substantial. Finite-cluster inference is somewhat more conservative but does not overturn the main conclusion. With a wild cluster bootstrap by release, the \(p\)-values are 0.018 for implied standard deviation, 0.033 for \(\Pr^K(y>0.4)\), 0.057 for \(\Pr^K(y>0.3)\), and 0.054 for \(\Pr^K(y>0.5)\). Thus the evidence is strongest for uncertainty and the middle fixed upper-tail threshold, with the other fixed tails remaining statistically meaningful under conservative inference. Consensus-relative tail probabilities remain weak.

The online appendix reports three further checks. First, alternative tail-support rules and a uniform-within-bin interpolation deliver positive fixed-tail coefficients across constructions, indicating that the tail results are not an artifact of assigning bin probabilities to a particular support point. Second, split-sample regressions for CPI and core CPI show the same qualitative pattern but less precision, as expected given the smaller number of release clusters. The uncertainty response is particularly strong for core CPI. Third, leave-one-release-out diagnostics show that the uncertainty and fixed-tail coefficients do not depend on a single release. The mean-gap coefficient is the least stable object, which is consistent with the paper's interpretation that the distinctive information in prediction-market distributions is not the conditional mean alone.

\begin{table}[h!]
\centering
\caption{Robustness summary for the main distributional regressions}
\label{tab:robustness_summary}
\scriptsize
\begin{tabular}{lccccc}
\toprule
Dependent variable & Baseline & Stale $\leq$ 1h & Controls & Wild $p$ & LOO range \\
\midrule
$\mu^K-C$ & -0.023 [0.828] & 0.173 [0.255] & -0.053 [0.618] & 0.825 & [-0.054, 0.046] \\
$\sigma^K$ & 0.136 [0.009] & 0.114 [0.038] & 0.121 [0.022] & 0.018 & [0.113, 0.164] \\
$\sigma^K$ + range & 0.089 [0.094] & 0.113 [0.048] & 0.084 [0.117] & 0.141 & [0.059, 0.120] \\
$\Pr^K(y>0.3)$ & 0.466 [0.042] & 1.198 [0.000] & 0.407 [0.069] & 0.057 & [0.329, 0.537] \\
$\Pr^K(y>0.4)$ & 0.422 [0.020] & 0.838 [0.035] & 0.356 [0.027] & 0.033 & [0.333, 0.488] \\
$\Pr^K(y>0.5)$ & 0.281 [0.039] & 0.441 [0.084] & 0.230 [0.079] & 0.054 & [0.237, 0.332] \\
\bottomrule
\end{tabular}
\vspace{0.5em}
\begin{minipage}{0.94\textwidth}
\footnotesize
\emph{Notes:} Entries in the first three coefficient columns report the coefficient on the lagged Reuters surprise, with finite-cluster \(p\)-values in brackets. The uncertainty specifications use the absolute lagged surprise; the third row additionally controls for the current Reuters forecast range. Fixed-tail regressions control for the current Reuters consensus. All specifications include variable-by-horizon fixed effects and cluster standard errors by release. The controls column adds snapshot staleness, \(\log(1+\mathrm{open\ interest})\), and \(\log(1+\mathrm{volume})\). The wild column reports two-sided wild cluster bootstrap \(p\)-values with 999 replications. LOO denotes leave-one-release-out.
\end{minipage}
\end{table}

\section{What Do Prediction-Market Beliefs Measure?}
\label{sec:discussion}

Prediction-market prices are attractive because their payoffs are directly linked to realized macroeconomic outcomes. This feature makes them closer to beliefs about specific statistical releases than many asset-price-based measures. At the same time, the recovered objects should not be interpreted mechanically as frictionless subjective probabilities. Under strong assumptions---risk neutrality, common priors, competitive trading, no fees, and sufficient liquidity---a binary contract price equals the probability of the payoff-relevant event. In practice, prediction-market prices can embed risk premia, heterogeneous beliefs, trading fees, bid-ask spreads, depth constraints, and stale quotes. Evidence from recent studies of Kalshi and Polymarket microstructure reinforces this point: event-contract prices are informative, but they can also reflect favorite-longshot bias, liquidity provision, platform design, and heterogeneous trader sophistication \citep{BurgiDengWhelan2026,Dubach2026,AkeyGregoireHarvieMartineau2026}. The objects in this paper are therefore market-implied belief distributions: they combine information, beliefs, risk preferences, and market microstructure.

This interpretation is analogous to other market-based expectation measures. Inflation swaps, TIPS breakevens, options, and futures provide valuable information, but they also embed risk and liquidity premia. Prediction markets have the advantage that the payoff is more directly tied to the macroeconomic outcome. Their disadvantage is that markets may be thinner and contract grids may be discrete or incomplete. The empirical design addresses these concerns by aggregating prices to hourly frequency, using standardized pre-release snapshots, clustering standard errors by release, and focusing on repeated patterns across events rather than on individual prices. The robustness exercises further show that the main uncertainty and fixed-tail results are not explained by stale quotes, open interest, volume, tail-support choices, or a single influential release. Survey density forecasts are an important benchmark but not a one-for-one substitute for the object studied here. The Survey of Professional Forecasters and related density data are designed for lower-frequency horizons and are well suited to studying disagreement, uncertainty, and macroeconomic risk over quarters or years \citep{Croushore1993,EngelbergManskiWilliams2009,AndradeCrumpEusepiMoench2016}. Kalshi contracts instead price the probability distribution of a specific release at intramonth horizons. The Reuters poll is therefore used as the main conditioning benchmark because it provides a release-level professional point forecast and a high-low range for the same CPI and core CPI announcements. The empirical claim is not that Kalshi replaces survey densities. It is that release-contingent prediction markets provide a complementary high-frequency distribution that is unavailable from standard point forecasts. The main economic interpretation is that recent inflation news affects perceived inflation risk more than disagreement between prediction markets and professional forecasters. The Kalshi mean does not systematically move above the Reuters consensus after a positive lagged surprise. Nor does the probability of exceeding the current consensus forecast respond strongly. Instead, fixed high-inflation tail probabilities rise. This pattern is consistent with a change in the perceived inflation-risk environment: the consensus forecast may remain a good summary of the center of the distribution, while the market assigns more probability to states that would be especially salient for monetary policy and financial markets.

\section{Conclusion}
\label{sec:conclusion}

Kalshi prediction-market contracts make it possible to recover high-frequency distributions of short-run macroeconomic beliefs. The empirical setting is CPI and core CPI releases, where adjacent threshold contracts imply probability mass over inflation outcomes at standardized horizons before release. The resulting data reveal not only market-implied means, but also uncertainty and upper-tail probabilities. The main result is distributional. Kalshi-implied means contain useful information about realized headline CPI and remain broadly aligned with Reuters Poll consensus forecasts. Lagged Reuters surprises do not predict systematic deviations of the Kalshi mean from the current Reuters consensus. By contrast, recent inflation news predicts changes in the shape of the distribution. Large previous surprises are associated with higher implied uncertainty, and positive previous surprises raise the probability assigned to fixed high-inflation outcomes even after controlling for the current consensus forecast. The robustness evidence supports this interpretation. The uncertainty and fixed-tail results survive staleness filters, continuous controls for staleness and market activity, alternative tail-support and interpolation rules, wild cluster bootstrap inference, and leave-one-release-out diagnostics. The strongest and most stable finding is not that prediction markets persistently move their mean above professional consensus after a surprise. It is that recent inflation news is reflected in market-implied inflation risk. These findings suggest that prediction markets can add to macroeconomic monitoring by measuring short-run inflation risk in real time. Point forecasts are useful, but they can miss changes in the distribution that matter for monetary policy and financial markets. A stable expected release can coexist with a higher probability of a high-inflation state. Prediction-market distributions make this distinction observable. Future work should extend the sample as these markets mature, compare prediction-market distributions with survey density forecasts and options-implied distributions, and use full pre-release distributions to construct distributional macroeconomic news measures for structural analysis.

\clearpage
\phantomsection
\addcontentsline{toc}{section}{Online Appendix}

\begin{center}
    {\Large \textbf{Online Appendix}}\\[0.4cm]
    {\large to}\\[0.25cm]
    {\LARGE \textbf{The Shape of Macroeconomic Beliefs}}\\[0.7cm]

    Giovanni Angelini \\
    {\small University of Bologna}\\[0.4cm]

    {\small June 2026}
\end{center}

\clearpage 

\appendix
\renewcommand{\thesection}{Appendix \Alph{section}}
\renewcommand{\thesubsection}{\Alph{section}.\arabic{subsection}}

\section{Moving-Average Benchmark Robustness}
\label{app:ma}

The main analysis uses Reuters Poll consensus forecasts to construct macroeconomic surprises. This appendix reports robustness exercises based on simple real-time moving-average benchmarks. The moving-average exercises are not the preferred identification strategy because moving averages are a weaker proxy for the information available to market participants. They are nevertheless useful because they are transparent, public, and do not rely on proprietary forecast data. Let \(MA3_r\) and \(MA6_r\) denote the three-month and six-month moving averages available before release \(r\). Define the lagged moving-average deviation as
\[
s^{MA3}_{r-1}=y_{r-1}-MA3_{r-1}.
\]
Horizon fixed-effect regressions relate market-implied means, uncertainty, and tail probabilities to lagged deviations from these benchmarks. Table~\ref{tab:recency} reports the mean and uncertainty results.

\begin{table}[!htbp]
\centering
\caption{Moving-average deviations and market-implied beliefs}
\label{tab:recency}
\begin{threeparttable}
\small
\begin{tabular}{lccccc}
\toprule
Dependent variable & Regressor & Obs. & Events & Coef. & s.e. \\
\midrule
\(\mu^K - MA3\)          & \(y_{r-1}-MA3_{r-1}\)       & 484 & 79 & 0.136 & 0.147 \\
\(\mu^K - MA6\)          & \(y_{r-1}-MA6_{r-1}\)       & 484 & 79 & 0.226 & 0.157 \\
\(\sigma^K\)             & \(|y_{r-1}-MA3_{r-1}|\)     & 484 & 79 & 0.122 & 0.032 \\
\(y_r-\mu^K\)            & \(y_{r-1}-MA3_{r-1}\)       & 484 & 79 & -0.079 & 0.088 \\
\bottomrule
\end{tabular}
\begin{tablenotes}
\footnotesize
\item \emph{Notes:} The table reports regressions relating market-implied beliefs to recent inflation deviations from moving-average benchmarks. The first two rows study the market-implied mean relative to MA3 and MA6 benchmarks. The third row studies the implied standard deviation. The fourth row asks whether the lagged deviation predicts the subsequent realized forecast error. All specifications include horizon fixed effects. Standard errors are clustered by release.
\end{tablenotes}
\end{threeparttable}
\end{table}

The moving-average results are consistent with the main Reuters-based findings. The response of the mean is positive but imprecise, while the implied standard deviation rises strongly after large lagged deviations. The coefficient on \(|y_{r-1}-MA3_{r-1}|\) is 0.122, with a clustered standard error of 0.032. Table~\ref{tab:tails} reports the corresponding upper-tail regressions. Positive lagged deviations increase the probability assigned to high-inflation outcomes. The coefficient for \(\Pr^K(y_r>0.3)\) is 0.332, with a standard error of 0.159. Tail probabilities relative to moving-average thresholds also respond positively, especially relative to the six-month moving average.

\begin{table}[!htbp]
\centering
\caption{Moving-average deviations and upper-tail beliefs}
\label{tab:tails}
\begin{threeparttable}
\small
\begin{tabular}{lccccc}
\toprule
Dependent variable & Regressor & Obs. & Events & Coef. & s.e. \\
\midrule
\(\Pr^K(y_r>0.3)\)   & \(y_{r-1}-MA3_{r-1}\) & 484 & 79 & 0.332 & 0.159 \\
\(\Pr^K(y_r>0.4)\)   & \(y_{r-1}-MA3_{r-1}\) & 484 & 79 & 0.240 & 0.180 \\
\(\Pr^K(y_r>MA3_r)\) & \(y_{r-1}-MA3_{r-1}\) & 484 & 79 & 0.147 & 0.154 \\
\(\Pr^K(y_r>MA6_r)\) & \(y_{r-1}-MA6_{r-1}\) & 484 & 79 & 0.403 & 0.174 \\
\bottomrule
\end{tabular}
\begin{tablenotes}
\footnotesize
\item \emph{Notes:} The table reports regressions of market-implied upper-tail probabilities on lagged inflation deviations from moving-average benchmarks. Tail probabilities are recovered from the market-implied distribution at each release-horizon observation. All specifications include horizon fixed effects. Standard errors are clustered by release.
\end{tablenotes}
\end{threeparttable}
\end{table}

\begin{figure}[!htbp]
\centering
\begin{subfigure}{0.48\textwidth}
\centering
\includegraphics[width=\textwidth]{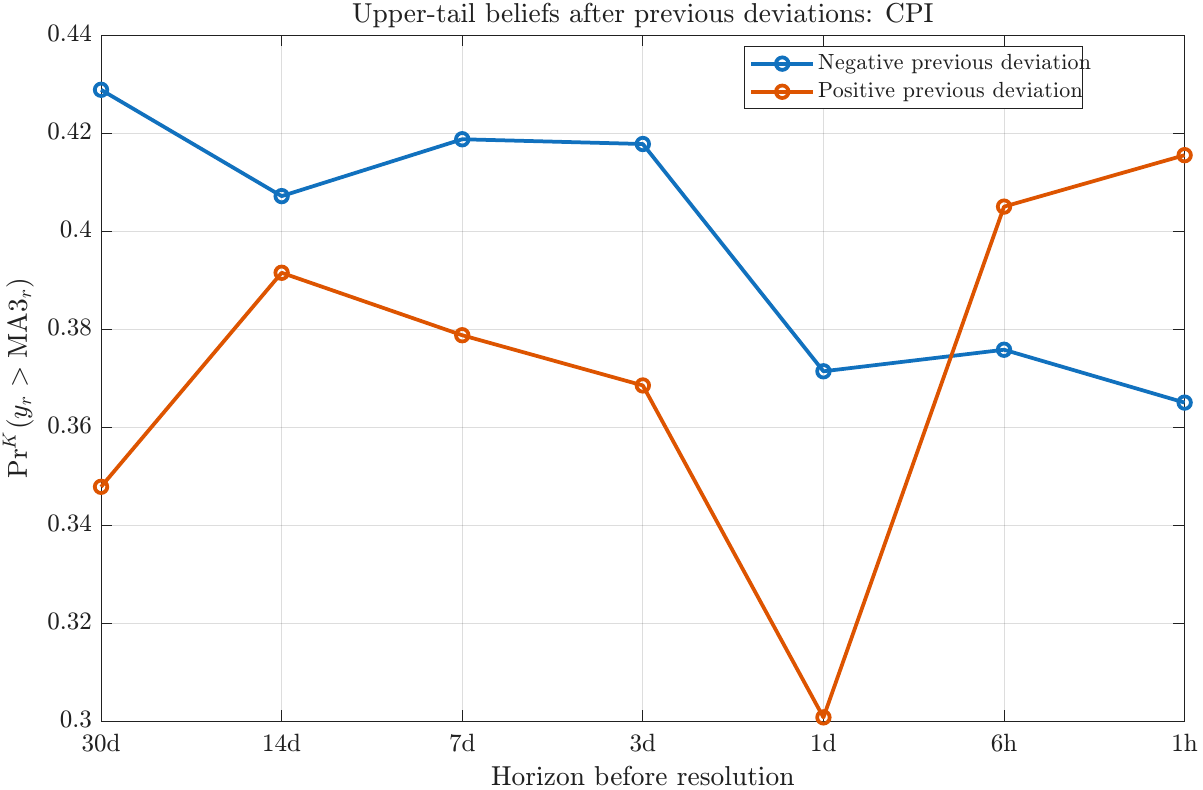}
\caption{CPI: tail above MA3}
\end{subfigure}
\hfill
\begin{subfigure}{0.48\textwidth}
\centering
\includegraphics[width=\textwidth]{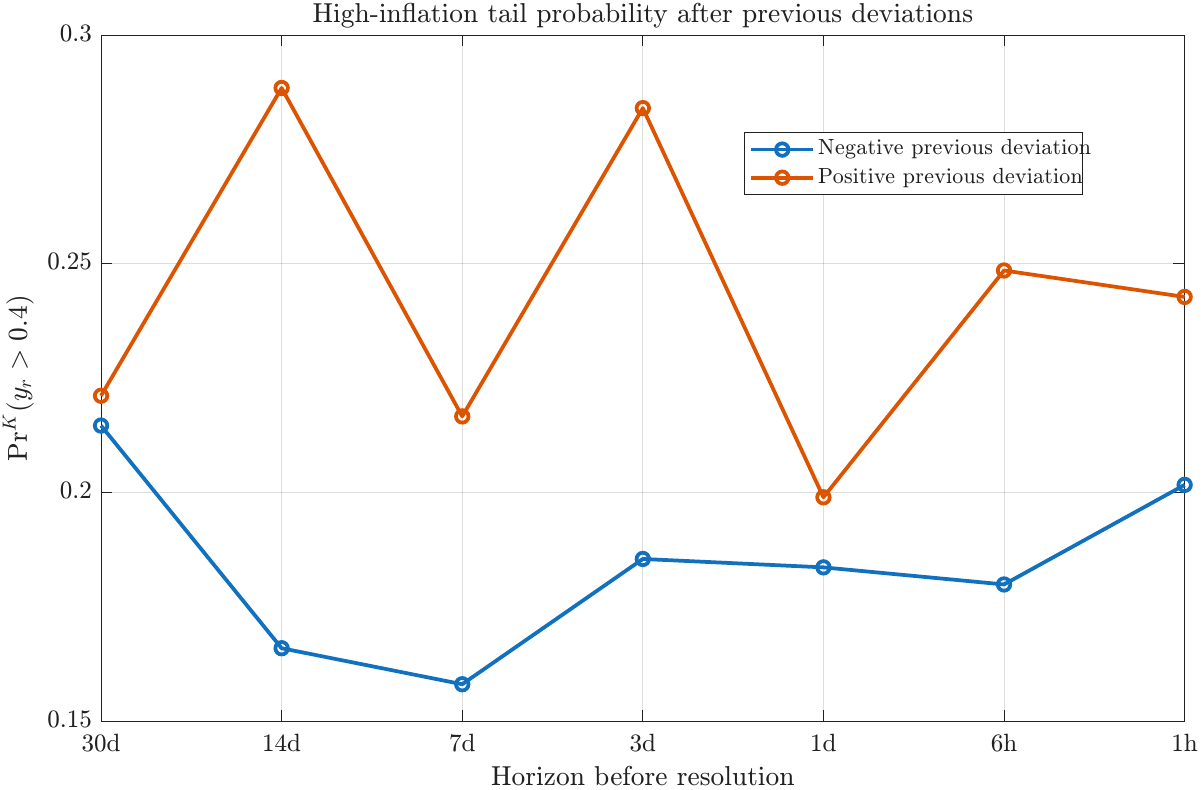}
\caption{High-inflation tail}
\end{subfigure}
\caption{Upper-tail beliefs after recent inflation deviations}
\label{fig:tail_event_study}
\begin{minipage}{0.9\textwidth}
\footnotesize
\emph{Notes:} Panel A reports the average market-implied probability that headline CPI exceeds its three-month moving-average benchmark, separately for releases following positive and negative lagged deviations. Panel B reports the average probability that inflation exceeds 0.4 percent month over month. Horizons are measured relative to market resolution.
\end{minipage}
\end{figure}

Figure~\ref{fig:tail_event_study} provides a graphical illustration for headline CPI. The probability assigned to high-inflation outcomes is generally higher after positive lagged deviations than after negative lagged deviations. These exercises are secondary to the Reuters consensus specifications, but they support the main conclusion that recent inflation news is reflected in distributional shape.

\section{Labor-Market External Validity}
\label{app:labor}

This appendix extends the consensus-based analysis to labor-market releases. The goal is not to make labor markets a second main empirical setting. The distributional mapping is less regular than for CPI, the relevant scales differ across variables, and the estimates are more sensitive to influential observations. The purpose is to test whether the inflation result has suggestive external validity in another salient macroeconomic domain. The labor-market sample combines Kalshi distributions with Reuters consensus forecasts for nonfarm payrolls and the unemployment rate. Table~\ref{tab:labor_coverage} reports the merged coverage. Nonfarm payroll forecasts and outcomes are measured in job levels. The original Reuters payroll series is reported in thousands and is multiplied by 1,000 before merging. Unemployment-rate variables are measured in percentage points.

\begin{table}[!htbp]
\centering
\caption{Labor-market prediction-market sample}
\label{tab:labor_coverage}
\begin{threeparttable}
\small
\begin{tabular}{lccccccc}
\toprule
Variable & Obs. & Events & Bins & Mean & Consensus & S.d. & Hours \\
\midrule
Nonfarm payrolls  & 245 & 35 & 5 & 164{,}000 & 170{,}000 & 63{,}744 & 10.07 \\
Unemployment rate & 299 & 44 & 4 & 3.869 & 3.900 & 0.120 & 11.92 \\
\bottomrule
\end{tabular}
\begin{tablenotes}
\footnotesize
\item \emph{Notes:} The table reports coverage of the merged Kalshi-Reuters labor-market sample. Hours measure the time between the target snapshot horizon and the most recent available market observation.
\end{tablenotes}
\end{threeparttable}
\end{table}

To put the two variables on a common directional scale, a positive labor-market surprise is defined as news indicating a stronger labor market. For nonfarm payrolls,
\[
s^L_r=NFP_r-C^{NFP}_r.
\]
For unemployment, stronger news corresponds to a lower-than-expected unemployment rate, so the sign is reversed:
\[
s^L_r=-(UR_r-C^{UR}_r).
\]
Strong-labor tails are defined relative to the Reuters consensus. For nonfarm payrolls, the strong-labor tail is \(\Pr^K(NFP_r>C^{NFP}_r+50{,}000)\). For unemployment, it is \(\Pr^K(UR_r<C^{UR}_r-0.1)\). Weak-labor tails are defined analogously on the opposite side of consensus. Because payrolls and unemployment are measured on different scales, the pooled specifications standardize surprises within variable group and include variable-by-horizon fixed effects. Table~\ref{tab:labor_standardized_robustness} reports the results. A one-standard-deviation positive labor-market surprise predicts a higher market-implied mean relative to consensus and a larger probability of a strong-labor tail. The strong-labor tail coefficient is 0.037 in the pooled sample, statistically significant at the one percent level. This coefficient remains positive when the largest surprises are trimmed, although precision weakens.

\begin{table}[!htbp]
\centering
\caption{External validity: labor-market releases}
\label{tab:labor_standardized_robustness}
\begin{threeparttable}
\small
\begin{tabular}{llrrrr}
\toprule
Sample & Dependent variable & Coef. & s.e. & \(p\)-value & Obs. \\
\midrule
Pooled & Mean gap, standardized & 0.119 & 0.055 & 0.032 & 530 \\
Pooled & Implied s.d., standardized & 0.180 & 0.144 & 0.211 & 530 \\
Pooled & Strong-labor tail & 0.037 & 0.014 & 0.009 & 530 \\
Pooled & Weak-labor tail & 0.017 & 0.012 & 0.156 & 530 \\
Pooled, drop top 5\% surprises & Strong-labor tail & 0.033 & 0.017 & 0.051 & 509 \\
Pooled, drop top 10\% surprises & Strong-labor tail & 0.033 & 0.018 & 0.068 & 503 \\
Nonfarm payrolls & Mean gap, standardized & 0.194 & 0.081 & 0.017 & 238 \\
Nonfarm payrolls & Implied s.d., standardized & 0.380 & 0.186 & 0.042 & 238 \\
Nonfarm payrolls & Strong-labor tail & 0.059 & 0.019 & 0.001 & 238 \\
Nonfarm payrolls & Weak-labor tail & 0.009 & 0.016 & 0.584 & 238 \\
Unemployment rate & Mean gap, standardized & 0.058 & 0.072 & 0.423 & 292 \\
Unemployment rate & Strong-labor tail & 0.019 & 0.020 & 0.346 & 292 \\
Unemployment rate & Weak-labor tail & 0.023 & 0.017 & 0.164 & 292 \\
\bottomrule
\end{tabular}
\begin{tablenotes}
\footnotesize
\item \emph{Notes:} The table reports standardized labor-market robustness regressions. A positive labor-market surprise denotes a stronger-than-expected release: a positive payroll surprise or a negative unemployment-rate surprise. Surprises and continuous dependent variables are standardized within variable group. For nonfarm payrolls, the strong-labor tail is \(\Pr^K(NFP>C+50{,}000)\); for unemployment, it is \(\Pr^K(UR<C-0.1)\). Weak-labor tails are defined analogously. Pooled specifications include variable-by-horizon fixed effects. Single-variable specifications include horizon fixed effects. Standard errors are clustered by release.
\end{tablenotes}
\end{threeparttable}
\end{table}

The pooled labor-market evidence is driven primarily by nonfarm payrolls. In the payroll sample, a one-standard-deviation positive lagged payroll surprise predicts a 0.194 standard-deviation increase in the market-implied mean relative to consensus, higher implied uncertainty, and a 5.9 percentage point increase in the probability of a strong-labor outcome. The unemployment-rate results are weaker and statistically insignificant. The labor-market analysis is therefore interpreted as suggestive external-validity evidence. It supports the broader idea that prediction markets can encode macroeconomic belief updating through distributional shape, but the main conclusions of the paper rest on the cleaner CPI and core CPI setting.

\section{Additional Robustness Diagnostics}
\label{app:robustness}

This appendix reports the main robustness diagnostics behind
the robustness section of the main manuscript. The goal is not to reproduce every possible specification, but to show that the central interpretation of the paper is not driven by market-quality filters, by the construction of tail probabilities, by finite-cluster inference, or by a small number of influential releases. Across the diagnostics below, the same pattern emerges. The coefficient on the Kalshi mean relative to the Reuters consensus is small and statistically insignificant in most specifications. By contrast, the coefficients on market-implied uncertainty and fixed upper-tail probabilities remain positive
across the main robustness checks. This is why the paper interprets the evidence as a change in perceived inflation risk, rather than as a robust shift in the conditional mean of the next release.

\subsection{Market quality: staleness and liquidity}

A first concern is that the estimates may be driven by stale prices or by thinly traded markets. Table~\ref{tab:app_market_quality} addresses this concern in three ways. The first two rows impose increasingly strict staleness filters, keeping only observations whose last market update is within six hours or one hour of the target snapshot. The next two rows impose open-interest filters. The last row includes continuous controls for quote staleness, open interest, and trading volume. The main distributional result is not weakened by these checks. If anything,
the coefficients on fixed upper-tail probabilities are often larger in the fresh-price samples. Open-interest filters also preserve the sign pattern, although precision declines as the sample is reduced. Volume filters are less informative because hourly volume is often zero even when open interest is positive.

\begin{table}[!htbp]
\centering
\caption{Market-quality robustness}
\label{tab:app_market_quality}
\begin{threeparttable}
\scriptsize
\resizebox{\textwidth}{!}{%
\begin{tabular}{lccccc}
\toprule
Specification
& $\mu^K-C$
& $\sigma^K$ + range
& $\Pr^K(y>0.3)$
& $\Pr^K(y>0.4)$
& $\Pr^K(y>0.5)$ \\
\midrule
Baseline
& -0.023 [0.828]
& 0.089 [0.094]
& 0.466 [0.042]
& 0.422 [0.020]
& 0.281 [0.039] \\

Stale $\leq$ 6h
& 0.091 [0.489]
& 0.094 [0.030]
& 0.910 [0.001]
& 0.616 [0.076]
& 0.368 [0.086] \\

Stale $\leq$ 1h
& 0.173 [0.255]
& 0.113 [0.048]
& 1.198 [0.000]
& 0.838 [0.035]
& 0.441 [0.084] \\

Positive open interest
& -0.020 [0.854]
& 0.091 [0.089]
& 0.474 [0.040]
& 0.433 [0.018]
& 0.292 [0.034] \\

Open interest $\geq$ p25
& -0.017 [0.886]
& 0.098 [0.103]
& 0.531 [0.022]
& 0.316 [0.066]
& 0.202 [0.136] \\

Staleness + log OI + log volume
& -0.053 [0.618]
& 0.084 [0.117]
& 0.407 [0.069]
& 0.356 [0.027]
& 0.230 [0.079] \\

\bottomrule
\end{tabular}}
\vspace{0.4em}
\begin{minipage}{0.94\textwidth}
\footnotesize
\emph{Notes:} Entries report coefficients with finite-cluster $p$-values in brackets. The staleness filter uses the time between the target snapshot and the most recent available market observation. OI denotes open interest. The final row adds continuous controls for snapshot staleness in hours, $\log(1+\mathrm{open\ interest})$, and $\log(1+\mathrm{volume})$. Uncertainty regressions use the absolute lagged Reuters surprise and control for the Reuters range. Fixed-tail regressions control for the current Reuters consensus. Standard errors are clustered by release.
\end{minipage}
\end{threeparttable}
\end{table}

\subsection{Construction of upper-tail probabilities}

A second concern is that the tail-risk results may depend on how mass inside Kalshi event-contract bins is mapped into survival probabilities. The baseline uses the midpoint rule. Table~\ref{tab:app_tail_construction} compares this rule with weak-threshold, lower-bound, upper-bound, and uniform-within-bin
constructions. The fixed upper-tail results are robust to these alternatives. The coefficient on $\Pr^K(y>0.3)$ is positive and statistically meaningful under all constructions. The coefficient on $\Pr^K(y>0.4)$ is also positive across all rules and remains statistically significant or marginally significant in each case. The result for $\Pr^K(y>0.5)$ is positive throughout, although the lower
conservative rule is less precise. This supports the interpretation that the main tail-risk result is not an artifact of a particular interpolation rule.

\begin{table}[!htbp]
\centering
\caption{Tail-construction robustness}
\label{tab:app_tail_construction}
\begin{threeparttable}
\scriptsize
\begin{tabular}{lccc}
\toprule
Tail construction
& $\Pr^K(y>0.3)$
& $\Pr^K(y>0.4)$
& $\Pr^K(y>0.5)$ \\
\midrule
Midpoint strict
& 0.466 [0.042]
& 0.422 [0.020]
& 0.281 [0.039] \\

Midpoint weak
& 0.571 [0.007]
& 0.487 [0.012]
& 0.324 [0.028] \\

Lower conservative
& 0.393 [0.031]
& 0.278 [0.039]
& 0.194 [0.165] \\

Upper liberal
& 0.551 [0.012]
& 0.579 [0.005]
& 0.400 [0.014] \\

Uniform within bin
& 0.492 [0.020]
& 0.471 [0.012]
& 0.317 [0.025] \\

\bottomrule
\end{tabular}
\vspace{0.4em}
\begin{minipage}{0.57\textwidth}
\footnotesize
\emph{Notes:} Entries report coefficients with finite-cluster $p$-values in brackets. The table varies the rule used to map contract-bin probabilities into fixed-threshold survival probabilities. All specifications control for the current Reuters consensus and include variable-by-horizon fixed effects. Standard errors are clustered by release.
\end{minipage}
\end{threeparttable}
\end{table}

\subsection{Finite-cluster inference}

The baseline inference clusters standard errors by release. Since the number of release clusters is moderate, Table~\ref{tab:app_wild_cluster} reports wild-cluster bootstrap $p$-values for the main outcomes.

The wild-cluster results are somewhat more conservative, as expected, but they do not change the substantive interpretation. The implied standard deviation remains statistically significant without the Reuters range and remains economically large with the range included. The fixed upper-tail coefficients remain positive, with bootstrap $p$-values close to or below conventional significance thresholds. The mean-gap coefficient remains small and insignificant.

\begin{table}[!htbp]
\centering
\caption{Wild-cluster bootstrap inference}
\label{tab:app_wild_cluster}
\begin{threeparttable}
\small
\begin{tabular}{lrrrrr}
\toprule
Dependent variable
& Coef.
& Cluster s.e.
& $p_{t(G-1)}$
& Wild $p$
& Obs. \\
\midrule
$\mu^K-C$
& -0.023
& 0.107
& 0.828
& 0.825
& 484 \\

$\sigma^K$
& 0.136
& 0.051
& 0.009
& 0.018
& 484 \\

$\sigma^K$ + Reuters range
& 0.089
& 0.053
& 0.094
& 0.141
& 484 \\

$\Pr^K(y>0.3)$
& 0.466
& 0.225
& 0.042
& 0.057
& 484 \\

$\Pr^K(y>0.4)$
& 0.422
& 0.178
& 0.020
& 0.033
& 484 \\

$\Pr^K(y>0.5)$
& 0.281
& 0.134
& 0.039
& 0.054
& 484 \\

\bottomrule
\end{tabular}
\vspace{0.4em}
\begin{minipage}{0.72\textwidth}
\footnotesize
\emph{Notes:} The table reports baseline coefficients with cluster-robust
standard errors by release and two-sided wild-cluster bootstrap $p$-values based on 999 replications. The uncertainty regressions use absolute lagged Reuters surprises. Fixed-tail regressions control for the current Reuters consensus. The specification with $\sigma^K$ plus Reuters range includes the Reuters high-low forecast range as an additional control.
\end{minipage}
\end{threeparttable}
\end{table}

\subsection{CPI/core CPI splits and influence diagnostics}

The final diagnostics ask whether the findings are concentrated in one inflation measure or in a small number of influential releases. These are demanding checks: estimating CPI and core CPI separately reduces the number of clusters, and leave-one-release-out regressions repeatedly discard an entire release event. Table~\ref{tab:app_split_influence} reports both exercises. Panel A estimates the main regressions separately for CPI and core CPI. The sign pattern is
similar across the two samples, but precision naturally falls. The uncertaintyresponse is particularly strong for core CPI, while fixed-tail coefficients are positive in both samples but less precisely estimated than in the pooled regressions. Panel B reports leave-one-release-out diagnostics. The uncertainty and fixed-tail coefficients remain positive in every leave-one-out sample. The mean-gap coefficient is the only main outcome that changes sign when individual
releases are excluded. Together, these checks reinforce the main interpretation. The robust part of the evidence is not a systematic movement of the Kalshi mean away from the Reuters consensus. It is the response of market-implied uncertainty and fixed upper-tail probability mass.

\begin{table}[!htbp]
\centering
\caption{Heterogeneity and influence diagnostics}
\label{tab:app_split_influence}
\begin{threeparttable}
\scriptsize
\resizebox{\textwidth}{!}{%
\begin{tabular}{llrrrrr}
\toprule
\multicolumn{7}{l}{\textbf{Panel A. Separate CPI and core CPI regressions}} \\
\midrule
Variable
& Dependent variable
& Obs.
& Clusters
& Coef.
& $p$-value
& Interpretation \\
\midrule
CPI
& $\mu^K-C$
& 262
& 43
& -0.110
& 0.500
& No robust mean shift \\

CPI
& $\sigma^K$ + Reuters range
& 262
& 43
& 0.067
& 0.337
& Positive, imprecise \\

CPI
& $\Pr^K(y>0.3)$
& 262
& 43
& 0.422
& 0.156
& Positive, imprecise \\

CPI
& $\Pr^K(y>0.4)$
& 262
& 43
& 0.340
& 0.167
& Positive, imprecise \\

CPI
& $\Pr^K(y>0.5)$
& 262
& 43
& 0.239
& 0.236
& Positive, imprecise \\

Core CPI
& $\mu^K-C$
& 222
& 36
& 0.115
& 0.154
& No robust mean shift \\

Core CPI
& $\sigma^K$ + Reuters range
& 222
& 36
& 0.137
& 0.042
& Stronger uncertainty response \\

Core CPI
& $\Pr^K(y>0.3)$
& 222
& 36
& 0.251
& 0.416
& Positive, imprecise \\

Core CPI
& $\Pr^K(y>0.4)$
& 222
& 36
& 0.404
& 0.179
& Positive, imprecise \\

Core CPI
& $\Pr^K(y>0.5)$
& 222
& 36
& 0.308
& 0.083
& Positive, marginal \\

\midrule
\multicolumn{7}{l}{\textbf{Panel B. Leave-one-release-out diagnostics}} \\
\midrule
Dependent variable
& Full coef.
& Min leave-one-out
& Max leave-one-out
& Largest change
& Sign flips
& Interpretation \\
\midrule
$\mu^K-C$
& -0.023
& -0.054
& 0.046
& 0.069
& 2
& Fragile mean gap \\

$\sigma^K$ + Reuters range
& 0.089
& 0.059
& 0.120
& 0.031
& 0
& Always positive \\

$\Pr^K(y>0.3)$
& 0.466
& 0.329
& 0.537
& 0.137
& 0
& Always positive \\

$\Pr^K(y>0.4)$
& 0.422
& 0.333
& 0.488
& 0.088
& 0
& Always positive \\

$\Pr^K(y>0.5)$
& 0.281
& 0.237
& 0.332
& 0.051
& 0
& Always positive \\

\bottomrule
\end{tabular}}
\vspace{0.4em}
\begin{minipage}{0.94\textwidth}
\footnotesize
\emph{Notes:} Panel A estimates the baseline regressions separately for CPI and core CPI. Standard errors are clustered by release. Panel B reports the distribution of coefficients obtained by dropping one release event at a time. The table reports only the main outcomes used in the paper: the mean gap, the implied standard deviation controlling for Reuters range, and fixed upper-tail
probabilities.
\end{minipage}
\end{threeparttable}
\end{table}
\section{Positioning Relative to Recent Prediction-Market Papers}
\label{app:related_prediction_markets}

A rapidly growing literature studies Kalshi and Polymarket as prediction-market platforms. Tables~\ref{tab:app_related_platforms_macro} and \ref{tab:app_related_platforms_market_quality} summarize the closest papers and clarify the margin on which this paper differs. These tables clarify the paper’s position relative to recent platform-specific work. The paper's main contribution is distributional rather than purely predictive. If prediction-market prices only provided another point forecast of the next CPI release, the incremental contribution relative to existing work would be limited. The distinctive information in the data is instead the sequence of market-implied probability distributions before each release. These distributions make it possible to study whether macroeconomic news changes the perceived location of inflation outcomes, the dispersion of beliefs, or the probability attached to high-inflation states. The robustness results in \ref{app:robustness} show that the central findings are strongest for uncertainty and fixed upper-tail probabilities, and are not driven by a single release, stale observations, or a particular tail-construction rule.

\begin{table}[h!]
\centering
\caption{Recent platform-specific papers: macro and expectations applications}
\label{tab:app_related_platforms_macro}
\begin{threeparttable}
\scriptsize
\resizebox{\textwidth}{!}{%
\begin{tabular}{p{0.22\textwidth}p{0.30\textwidth}p{0.26\textwidth}p{0.18\textwidth}}
\toprule
Paper & Main object & Relation to this paper & Key distinction \\
\midrule
\citet{DiercksKatzWright2026} & Kalshi macro markets as real-time forecasts for macroeconomic releases. & Closest paper on Kalshi macro expectations. & This analysis studies distributional inflation beliefs and tail-risk updating, rather than mainly forecast accuracy across macro releases. \\
\citet{SwansonWangWu2025} & Effects of monetary-policy announcements on Kalshi-implied macroeconomic expectations. & Shares the use of Kalshi macro contracts for high-frequency expectations. & This analysis studies pre-release inflation distributions and lagged macro-news updating, not FOMC-event responses. \\
\citet{EichengreenViswanathNatrajWangWang2025} & Polymarket beliefs about Fed Chair removal and monetary-policy outcomes. & Shows that prediction markets can measure monetary-policy beliefs and institutional-credibility concerns. & This analysis uses release-contingent macro contracts to measure inflation risk rather than political-pressure beliefs. \\
\citet{GomezCramGuoJensenKung2025Earnings} and \citet{RabettiShaoZhang2026} & Prediction-market measures of earnings expectations. & Shows that prediction markets can generate high-frequency, financially backed expectation measures outside macro releases. & This analysis studies macroeconomic statistical releases and the shape of inflation beliefs. \\
\citet{DudleyMagdaleno2026} & Prediction-market performance in infectious-disease forecasting. & Provides a cautionary counterexample: prediction markets need not dominate simple benchmarks in every domain. & The present analysis shows forecast content but centers the contribution on distributional inflation risk. \\
\bottomrule
\end{tabular}}
\vspace{0.4em}
\begin{minipage}{0.94\textwidth}
\footnotesize
\emph{Notes:} The table focuses on recent platform-specific work with direct links to expectations, macroeconomic monitoring, or forecast evaluation. Some papers are recent working papers or preprints; details should be rechecked before submission.
\end{minipage}
\end{threeparttable}
\end{table}

\clearpage

\begin{table}[h!]
\centering
\caption{Recent platform-specific papers: market quality, microstructure, and traders}
\label{tab:app_related_platforms_market_quality}
\begin{threeparttable}
\scriptsize
\resizebox{\textwidth}{!}{%
\begin{tabular}{p{0.22\textwidth}p{0.30\textwidth}p{0.26\textwidth}p{0.18\textwidth}}
\toprule
Paper & Main object & Relation to this paper & Key distinction \\
\midrule
\citet{BurgiDengWhelan2026} & Kalshi pricing, maker-taker microstructure, favorite-longshot bias, and returns. & Motivates caution in interpreting prices as frictionless probabilities. & This analysis uses repeated macro distributions and controls for liquidity/staleness rather than modeling platform-wide trading returns. \\
\citet{TsangYang2026} and \citet{TsangYang2026PoliticalShocks} & Polymarket election-market transactions and political-shock price discovery. & Relevant for prediction-market price discovery and event responses. & The present analysis focuses on regulated macro event contracts and distributional expectations rather than election outcomes. \\
\citet{Dubach2026} & Polymarket order-book microstructure and measurement problems in trade direction. & Relevant for market-quality and measurement concerns. & This analysis works with event-contract prices and distributional objects, and addresses staleness and liquidity directly. \\
\citet{AkeyGregoireHarvieMartineau2026} and \citet{GomezCramGuoJensenKung2026Accuracy} & Trader-level profitability and the concentration of price discovery on Polymarket. & Reinforces that prices may reflect informed minorities and heterogeneous trader sophistication. & This analysis studies the aggregate belief distribution embedded in macro contracts, not who earns trading profits. \\
\citet{SaguilloGhafouriKifferSuarezTangil2025} & Arbitrage and cross-market probability consistency on Polymarket. & Relevant for interpreting probability prices and market efficiency. & This analysis recovers within-release distributions from adjacent macro thresholds and checks tail-support robustness. \\
\citet{MittsOfir2026} & Informed trading and regulatory issues in prediction markets. & Highlights information and manipulation risks in event markets. & This analysis uses repeated macro releases and robustness diagnostics rather than idiosyncratic political or entertainment events. \\
\bottomrule
\end{tabular}}
\vspace{0.4em}
\begin{minipage}{0.94\textwidth}
\footnotesize
\emph{Notes:} The table focuses on market-quality and trader-composition papers because these papers inform how prediction-market prices should be interpreted. The broader prediction-market literature is discussed in the introduction.
\end{minipage}
\end{threeparttable}
\end{table}

\end{document}